%% file: root.tex
\documentclass[letterpaper, 10 pt, conference]{ieeeconf}  % Comment this line out if you need a4paper

\IEEEoverridecommandlockouts                              % This command is only needed if 
\title{\LARGE \bf
A Method for Reducing the Complexity of Model Predictive Control in Robotics Applications
}

\author{Michael Muehlebach and Raffaello D'Andrea$^{1}$ % <-this % stops a space
\thanks{$^{1}$Both authors are with the Institute for Dynamic Systems and Control, ETH Zurich, Switzerland. The contact author is Michael Muehlebach, {\tt\small michaelm@berkeley.edu}. This work was supported by ETHGrant ETH-48 15-1.}%
}

\let\labelindent\relax

\newif\ifArxiv
\Arxivtrue % Arxiv version
\input{michael-preamble}

\input{def}

\begin{document}

\maketitle
\thispagestyle{empty}
\pagestyle{empty}

%%%%%%%%%%%%%%%%%%%%%%%%%%%%%%%%%%%%%%%%%%%%%%%%%%%%%%%%%%%%%%%%%%%%%%%%%%%%%%%%
\begin{abstract}
This article describes an approach for parametrizing input and state trajectories in model predictive control. The parametrization is designed to be invariant to time shifts, which enables warm-starting the successive optimization problems and reduces the computational complexity of the online optimization. It is shown that in certain cases (e.g. for linear time-invariant dynamics with input and state constraints) the parametrization leads to inherent stability and recursive feasibility guarantees without additional terminal set constraints. Due to the fact that the number of decision variables are greatly reduced through the parametrization, while the warm-starting capabilities are preserved, the approach is suitable for applications where the available computational resources (memory and CPU-power) are limited.
\end{abstract}

%%%%%%%%%%%%%%%%%%%%%%%%%%%%%%%%%%%%%%%%%%%%%%%%%%%%%%%%%%%%%%%%%%%%%%%%%%%%%%%%

\input{intro}
\input{problemFormulation}
\input{dynamics}
\input{constraints}

\input{resultingOpti}
\input{stability}
\input{simResults}
\input{expResults}
\input{conclusion}

\bibliographystyle{IEEEtran}
\bibliography{IEEEabrv,literature}

\ifArxiv
%\newpage
\appendices
\input{appendixNew}
\else
\fi

\end{document}

%% file: michael-preamble.tex
\usepackage[american]{babel}
\usepackage{hyperref}
\usepackage{epsfig} % for postscript graphics files
\usepackage{times} % assumes new font selection scheme installed
\usepackage{amsmath}
\usepackage{amsfonts}
\usepackage{amssymb}  % assumes amsmath package installed
\usepackage{tabularx}

\usepackage{xcolor}
\usepackage{transparent}
\usepackage{graphicx}
\graphicspath{{media/}}
\usepackage{tikz}
\usetikzlibrary{shapes,arrows}
\usepackage{epstopdf}
\usepackage{units}
\usepackage{pgf,pgfplots}
\usepackage[]{algorithm2e}
\usepackage{enumitem} % for adjusting margin in itemize

\usepackage{pstool}

\graphicspath{{media/}}
\newcommand{\executeiffilenewer}[3]{%
 \ifnum\pdfstrcmp{\pdffilemoddate{#1}}%
 {\pdffilemoddate{#2}}>0%
 {\immediate\write18{#3}}\fi%
}

\usepackage{arydshln}
\usepackage{cleveref}

\newcommand{\HChange}[1]{{#1}}
\newtheorem{theorem}{Theorem}[section]

\newtheorem{proposition}[theorem]{Proposition}

\newcommand{\qed}{\nobreak \ifvmode \relax \else
      \ifdim\lastskip<1.5em \hskip-\lastskip
      \hskip1.5em plus0em minus0.5em \fi \nobreak
      \vrule height0.5em width0.5em depth0.25em\fi}
      

%% file: def.tex
\newcommand{\TT}{\ensuremath{\mathsf{\tiny{T}}}}
\newcommand{\T}{^{\TT}}

%% file: intro.tex
\section{Introduction}
Model predictive control (MPC) has become a well-known
and widely used control strategy for solving challenging control
problems. Unlike many other approaches, MPC addresses
input and state constraints in a systematic way. It is based
on repeatedly solving an optimal control problem, including
the actual state as an initial condition and a prediction of
the system's evolution. This leads naturally to an implicit
feedback law, providing robustness against modeling errors
and disturbances, \cite{morari1999model}.

Model predictive control has been successfully applied to many robotic systems: For example, the authors of \cite{steering} present an MPC-based steering controller for an autonomous vehicle. The controller is shown to perform complex steering maneuvers in an emergency scenario that includes a double lane change on snow. A real-time path planner for an all-terrain vehicle that is relying on MPC is presented in \cite{all-terrain}, \HChange{the MPC-based control architecture of an autonomous wheelchair is discussed in \cite{wheelchair}}, and MPC is used to control an unmanned rotorcraft in \cite{rotorcraft}.

In many applications, however, computational resources are limited, and therefore it is often desirable to simplify the resulting optimization problems that are solved at every time step. This can be done, for example, by exploiting the specific system's structure, \cite{MarkMPC}, or by a careful design of the control structure, \cite{MahonyMPC}. In this work we propose a different strategy, that is based on systematically reducing the degrees of freedom by parametrizing input and state trajectories with basis functions. We will show that the proposed parametrization evolves naturally from the requirement that the resulting trajectories should be invariant to time shifts, meaning that a previously calculated input and/or state trajectory can be used to warm start the optimization at the next time step. 

In addition, it will be argued that by choosing exponentially decaying basis functions the truncation of the prediction horizon can be avoided, which, combined with the time-shift property, leads to straightforward recursive feasibility and closed-loop stability guarantees.

\emph{\HChange{Related works}:} Several strategies have been suggested to reduce the complexity of MPC. For instance, move-blocking describes the general idea of fixing the input or its derivative to be constant over several time steps, \cite{moveBlockingGondhalekar}. Many different flavors and variants exist, ranging from simply clamping certain inputs together to more elaborate, time-dependent strategies, such as moving window blocking, that are designed to preserve closed-loop stability and recursive feasibility guarantees, \cite{moveBlockingMorari}. A different point of view is \HChange{adopted} in \cite{WangMPC}, \HChange{\cite{ExploitingRossiterJournal}}, and \cite{AlternativeRossiter}, for example, where the input is parametrized with Laguerre and/or Kauz basis functions.

In \cite{WaveletMPC}, multiresolution analysis is used for parametrizing the input trajectory. However, the approach is mainly applicable to open-loop stable systems, where the impulse response is assumed to be negligible after a certain time horizon. For dealing with unstable systems, the proposed approach would require additional terminal constraints on the unstable modes. Similarly, the authors from \cite{SummersMultiscale} apply the wavelet transformation for simplifying the control laws obtained with explicit model predictive control. They show that the resulting simplified control law is everywhere feasible and quantify the suboptimality.

Our contribution is twofold: We show that requiring the parametrization to be invariant with respect to time-shifts imposes a specific structure on the basis functions used for representing the input (and/or state) trajectories. We will highlight that the proposed parametrization captures the classical input description, where no restriction on the input sequence is imposed as a special case. Likewise, our parametrization can be chosen such that Laguerre functions are obtained as a special case, as for example described in \cite{ExploitingRossiter}. In addition, we show that choosing the basis functions to be decaying, an infinite prediction horizon can be retained, yielding inherent closed-loop stability and recursive feasibility guarantees.

The input and state parametrizations that will be introduced in the following can be viewed as approximations to the underlying constrained linear quadratic regulator problem. This point of view has been explored in the technical report \cite{muehlebachTechnicalReport}, where various approximation results (including convergence) are discussed. Preliminary results appeared in the conference papers \cite{muehlebachParametrized} and \cite{muehlebachCDC}. In \cite{ImplementionMatthias}, and \cite{Implementation} a similar approach has been applied to control a flying vehicle actuated by ducted fans. \HChange{Unlike the earlier contributions, this article presents a general and flexible parametrization strategy for reducing the computational complexity of MPC, when controlling nonlinear systems. It adopts a discrete-time point of view, and presents previous stability and recursive feasibility results in a unified way. The discrete-time formulation enables an exact description of the sample-and-hold process, which is not possible with the earlier continuous-time results. The effectiveness of the approach is demonstrated by performing a swing up of the inverted pendulum-on-a-cart system, which requires, unlike earlier applications, to fully account for the nonlinear dynamics. In addition, a principled way of choosing the basis functions that parametrize input and state trajectories is discussed, and a comparison to a different parametrization approach presented in the literature (\cite[Ch.~5]{WangMPC}) is presented. The comparison is in terms of execution time and closed-loop performance. We also comment on the feasibility of the resulting optimization problems.}

\emph{Outline:} Sec.~\ref{Sec:ProbForm} introduces the problem formulation and the notation that is used throughout the article. The parametrization of the input and state trajectories is discussed in Sec.~\ref{Sec:Param}. Sec.~\ref{Sec:Dyn} covers the representation of the dynamics, whereas Sec.~\ref{Sec:Constr} shows how to include input and state constraints. The resulting optimization problems are summarized in Sec.~\ref{Sec:ResOpti}, and the stability results are included in Sec.~\ref{Sec:Stability}. \HChange{The approach is compared to \cite[Ch.~3]{WangMPC} in Sec.~\ref{Sec:Sim} and the application to a nonlinear real-world system is shown in Sec.~\ref{Sec:Exp}.} The article concludes with a summary in Sec.~\ref{Sec:Concl}.

%% file: problemFormulation.tex
\section{Problem formulation}\label{Sec:ProbForm}
We consider the following optimization problem as a starting point
\begin{align}
\inf &\sum_{k=0}^{\infty} l(k,x(k),u(k)) \label{eq:Prob}\\
\text{s.t. } &x(k+1)=f(k,x(k),u(k)), \nonumber\\
&g(x(k),u(k)) \leq 0, \quad k\in \mathbb{Z}^+,\nonumber\\
&x(0)=x_0, \nonumber
\end{align}
where the function $l: \mathbb{Z}^+ \times \mathbb{R}^n \times \mathbb{R}^m \rightarrow \mathbb{R}$ describes the running cost, $g: \mathbb{R}^n \times \mathbb{R}^m \rightarrow \mathbb{R}^{n_\text{c}}$ the constraints, and $f: \mathbb{Z}^+ \times \mathbb{R}^n \times \mathbb{R}^m \rightarrow \mathbb{R}^n$ the dynamics. The set of all nonnegative integers is denoted by $\mathbb{Z}^+$, the real numbers by $\mathbb{R}$. The integer $n$ describes the state dimension, $m$ the input dimension, and $n_\text{c}$ the number of constraints. The initial condition is denoted by $x_0$. The above formulation encodes finite horizon problems as a special case, as the running cost can be chosen to be zero for all time indices larger than the given horizon.

\section{The parametrization}\label{Sec:Param}
We will parametrize the input and state trajectory using basis functions, that is,
\begin{align}
\tilde{x}(k)=(I_n \otimes \tau(k))\T \eta_x, \quad \tilde{u}(k)=(I_m \otimes \tau(k))\T \eta_u,
\end{align} 
where $\otimes$ refers to the Kronecker product, and $\eta_x \in \mathbb{R}^{ns}$ and $\eta_u \in \mathbb{R}^{ms}$ to the parameter vectors. The basis functions are captured with the vectors $\tau(k) \in \mathbb{R}^s$, $k\in \mathbb{Z}^+$, \HChange{where the integer $s>0$ refers to the number of basis functions.} We use the same basis functions to describe input and state trajectories, but this not necessarily needs to be the case. The choice is motivated by the fact that, as a result, linear time-invariant dynamics can be expressed by a very simple linear relationship between the parameters $\eta_x$ and $\eta_u$. Throughout the article we will denote parametrized trajectories with a tilde. We will make the following fundamental assumption:

\textbf{Assumption}
A1) The basis functions $\tau(k)$, $k\in \mathbb{Z}^+$ are linearly independent.

This assumption is necessary for the solutions in \eqref{eq:Prob}, when $x(k)$ and $u(k)$ are replaced with $\tilde{x}(k)$ and $\tilde{u}(k)$, to be unique (provided that they exist).

In MPC, \eqref{eq:Prob} will be optimized (up to a given tolerance) at every time step, subject to changing initial conditions $x_0$. In order to facilitate computation, we would like to warmstart the optimization at the next time step with the trajectories that were computed at the previous time step. This implies that for any $\eta \in \mathbb{R}^s$, there exists a parameter vector $\hat{\eta}(\eta) \in \mathbb{R}^s$, such that
\begin{equation}
\tau(k+1)\T \eta = \tau(k)\T \hat{\eta}(\eta), \quad \forall k\in \mathbb{Z}^+. \label{eq:timeshift}
\end{equation}
Due to the fact that the basis functions are linearly independent there are $s$ time indices $I:=\{k_1, k_2, \dots k_s\}$ such that the matrix
\begin{equation}
T:=\left[ \tau(k)\T \right]_{k\in I} = \left( \begin{array}{c} \tau(k_1)\T \\ \tau(k_2)\T \\ \vdots \end{array} \right) \label{eq:Tfr}
\end{equation}
is full rank. Thus, \eqref{eq:timeshift} implies that
\begin{equation}
\left[\tau(k+1)\T\right]_{k\in I} \eta= T \hat{\eta}(\eta),
\end{equation}
which concludes that $\hat{\eta}$ is a linear function of $\eta$ (multiplication of both sides with $T^{-1}$). Combined with the fact that \eqref{eq:timeshift} is required to hold for all $\eta\in \mathbb{R}^s$, we obtain, by taking the derivative of \eqref{eq:timeshift} with respect to $\eta$,
\begin{equation}
\tau(k+1) = M \tau(k), \quad \forall k \in \mathbb{Z}^+, \label{eq:bftmp}
\end{equation}
where the constant matrix $M\in \mathbb{R}^{s\times s}$ contains the partial derivative of $\hat{\eta}$ with respect to $\eta$. This concludes that if the basis function are required to fulfill relation \eqref{eq:timeshift}, they will automatically satisfy \eqref{eq:bftmp} for some matrix $M \in \mathbb{R}^{s\times s}$.

Conversely, if the basis functions satisfy \eqref{eq:bftmp}, for some matrix $M \in \mathbb{R}^{s\times s}$, \eqref{eq:timeshift} is satisfied for all $\eta \in \mathbb{R}^s$ with $\hat{\eta}=M\T \eta$.
The result is summarized by the following proposition.
\begin{proposition}
The basis functions satisfy \eqref{eq:bftmp} for some matrix $M \in \mathbb{R}^{s \times s}$ if and only if they are invariant under time shifts in the sense of \eqref{eq:timeshift}.
\end{proposition}

In the following we will focus on the regulation problem, and without loss of generality we assume that $x=0$, $u=0$ is an equilibrium of the dynamics. It is thus sensible to require that the basis functions should eventually decay to zero, leading to the requirement 
that \HChange{the dynamic system \eqref{eq:bftmp} should be asymptotically stable, i.e. the matrix $M$ should have all eigenvalues strictly within the unit circle.} Moreover, we assume that the function $g$ describing the constraints is continuous and satisfies $g(0,0)< 0$. The case where the equilibrium lies right on the boundary of the constraint manifold has limited practical importance, since, due to measurement noise, the resulting MPC algorithm is likely to become infeasible, even for initial conditions that are in a neighborhood of the equilibrium. The regularity assumptions on $g$ are needed for the closed-loop stability and recursive feasibility statements in Sec.~\ref{Sec:Stability}, where they guarantee existence of the minimizer of \eqref{eq:Prob}, for example.

Summarizing, we thus impose the following requirements on the basis functions
\begin{itemize}
\item [A1)] The basis functions $\tau(k), k\in \mathbb{Z}^+$ are linearly independent.
\item [A2)] The basis functions $\tau(k), k\in \mathbb{Z}^+$ satisfy the relation \eqref{eq:bftmp} for some matrix $M\in \mathbb{R}^{s\times s}$ that has all its eigenvalues within the unit circle.
\end{itemize}

\subsection{Examples}
There are numerous examples of basis functions that comply with Assumptions A1 and A2. 
\begin{itemize}[leftmargin=*]
\item The choice 
\begin{equation}
M=\left( \begin{array}{cccc} 0 &0 &0 &\hdots \\ 1 &0 &0 &\hdots \\ 0 &1 &0 &\hdots\\ \vdots & \vdots & \vdots & \ddots \end{array} \right), \quad \tau(0)=(1,0,0,\dots)\T, \label{eq:classicChoice}
\end{equation}
yields the classical approach, c.f. \cite{morari1999model}, where, in this case, the first $s$ inputs can be chosen arbitrarily. More precisely, if $\tilde{u}(k)=\tau(k)\T \eta$, then $\tilde{u}(0)=\eta_1$, $\tilde{u}(1)=\eta_2$, and so on, until $\tilde{u}(s)=0$, where $\eta=(\eta_1, \eta_2, \dots, \eta_s)\T$.
\item The choice
\begin{equation}
M=e^{M_\text{c} T_\text{s}}, \quad M_\text{c}=\left( \begin{array}{cccc} -\nu & 0 &0 & \hdots \\ -2\nu &-\nu & 0 & \hdots \\ -2\nu &-2\nu &-\nu & \hdots \\ \vdots & \vdots &\vdots &\ddots \end{array} \right), \label{eq:Laguerre}
\end{equation}
$\tau(0)=\sqrt{2 \nu}~(1,1,\dots)\T$, yields so-called Laguerre functions, \cite{LongHorizonRossiter}. These are essentially exponentially decaying polynomials, in the following sense
\begin{equation}
\tau(k) \in e^{-\nu k T_\text{s}} \text{span}( 1, k T_\text{s}, (k T_\text{s})^2, \dots),
\end{equation}
where $\nu$ is the decay rate ($\text{s}^{-1}$) and $T_\text{s}$ a  sampling time.
\item In an analogous way, the basis functions could be chosen such that 
\begin{align}
\tau(k) \in e^{-\nu k T_\text{s}} &\text{span}(1, \sin(\omega k T_\text{s}), \cos(\omega k T_\text{s}), \nonumber\\
&\quad\sin(2 \omega k T_\text{s}), \cos(2 \omega k T_\text{s}), \dots),
\end{align}
where $\omega$ denotes the frequency ($\text{rad}/\text{s}$).
\item In case the dynamics in \eqref{eq:Prob} are linear time-invariant, the cost is quadratic, and constraints are absent ($g=0$), the choice $M=A+BK$, where $A$ and $B$ refer to the system dynamics, and $K$ to the infinite-horizon linear quadratic regulator gain, recovers the solutions to the infinite-horizon linear quadratic regulator problem (see \cite{BFOpti} for the continuous-time analogue).
\end{itemize}
Any superposition of the above choices is valid as well, and is obtained by a blockdiagonal choice of the matrix $M$. \HChange{In Sec.~VIII we present a systematic choice for Laguerre basis functions (according to \eqref{eq:Laguerre}) based on optimizing the closed-loop performance.}

%% file: dynamics.tex
\section{Dynamics}\label{Sec:Dyn}
We propose to encode the dynamics using a variational formulation. We first note that the dynamics given in \eqref{eq:Prob} can be restated as
\begin{equation}
\sum_{k=0}^{\infty} \delta p(k)\T (x(k+1)-f(k,x(k),u(k))) = 0,
\end{equation}
for all variations $\delta p(k) \in \mathbb{R}^n$, $k\in \mathbb{Z}^+$. We apply the Galerkin approach to obtain a nonlinear equality constraint that approximates the dynamics. More precisely, we require the variations to be spanned by the basis functions and insert the parametrizations for the input and state trajectories. This yields
\begin{equation*}
\sum_{k=0}^{\infty} \delta \eta_p\T (I_n \otimes \tau(k)) (\tilde{x}(k+1) - f(k, \tilde{x}(k), \tilde{u}(k))= 0,
\end{equation*}
for all vectors $\delta \eta_p \in \mathbb{R}^{ns}$, or equivalently
\begin{equation}
\sum_{k=0}^{\infty} (I_n \otimes \tau(k)) (\tilde{x}(k+1)-f(k,\tilde{x}(k), \tilde{u}(k)))=0. \label{eq:dynTmp}
\end{equation}
Note that the above equation defines a nonlinear equality constraint between the parameter vectors $\eta_x$ and $\eta_u$ that parametrize input and state trajectories. If $f$ is continuous about $x=0$, $u=0$ uniformly in $k$, then the above sum is guaranteed to exist, due to the exponential decay of the basis functions.

For gaining additional insights we will now highlight two special cases.

\subsection{Linear time-invariant dynamics}
In this case $f(k,x(k),u(k))=A x(k) + B u(k)$, where $A$ and $B$ are real matrices of appropriate size. As a result, \eqref{eq:dynTmp} reduces to
\begin{align}
(I_n \otimes &\sum_{k=0}^{\infty} \tau(k)\tau(k)\T) \nonumber\\ &((I_n \otimes M\T - A \otimes I_s)\eta_x - (B \otimes I_s) \eta_u)=0,
\end{align}
where the property \eqref{eq:bftmp} of the basis functions and the properties of the Kronecker product have been exploited.
By the linear independence of the basis functions, the matrix $I_n \otimes \sum_{k=0}^{\infty} \tau(k)\tau(k)\T$ has full rank,\footnote{The matrix is positive definite, as shown by a contradiction argument: If there existed a vector $\eta\neq0$ such that $\eta\T \sum_{k=0}^{\infty} \tau(k)\tau(k)\T \eta =0$, this would imply that $\tau(k)\T \eta=0$ $\forall k \in \mathbb{Z}^+$, contradicting the fact that the basis functions are linearly independent.} and thus the above equation can be restated as
\begin{equation}
(I_n \otimes M\T - A \otimes I_s) \eta_x - (B \otimes I_s) \eta_u = 0. \label{eq:DynConst}
\end{equation}
The linear constraint \eqref{eq:DynConst} implies that the state and input trajectories fulfill
\begin{equation}
\tilde{x}(k+1)=A \tilde{x}(k) + B \tilde{u}(k), \quad \forall k \in \mathbb{Z}^+, \label{eq:linDyn}
\end{equation}
that is, the dynamics are fulfilled exactly (as can be seen by multiplying \eqref{eq:DynConst} with $(I_n \otimes \tau(k)\T)$ from the left). Linear independence of the basis functions can be used to conclude the converse. We will not cover the details, since the argument parallels Sec.~\ref{Sec:Param}, hinging on the fact that the matrix \eqref{eq:Tfr} is full rank. The result is summarized by the following proposition.
\begin{proposition}\label{Prop:LinDynamics}
Provided that the basis functions fulfill Assumptions A1 and A2, and that the dynamics are linear time-invariant, \eqref{eq:DynConst} and \eqref{eq:linDyn} are equivalent.
\end{proposition}

\subsection{Basis functions chosen according to \eqref{eq:classicChoice}}
In case the basis functions are chosen according to \eqref{eq:classicChoice}, the first $s$ variations are fully decoupled. As result, \eqref{eq:dynTmp} reduces to 
\begin{equation}
\tilde{x}(k+1)=f(k,\tilde{x}(k),\tilde{u}(k)), \quad k\in \{0,1,\dots,s\},
\end{equation}
that is, the dynamics are fulfilled exactly for the first $s$ time steps.

%% file: constraints.tex
\section{Constraints}\label{Sec:Constr}
The following section discusses the implementation of the inequality constraints,
\begin{equation}
g(\tilde{x}(k),\tilde{u}(k))\leq 0, \quad \forall k \in \mathbb{Z}^+. \label{eq:constr}
\end{equation}
It turns out that in many cases the constraint needs only to be checked at finitely many time instances, due to the decaying nature of the basis functions. We will relate the implementation of the inequality constraints to the well-studied problem of computing the maximum output admissible set for an autonomous linear time-invariant system. To that extent we define the variable $\tilde{z}(k)$ as 
\begin{equation}
\tilde{z}(k):=(\tilde{x}(k), \tilde{u}(k))=(I_{n+m} \otimes \tau(k))\T \eta_z,
\end{equation}
where $\eta_z:=(\eta_x,\eta_u)$ is obtained by stacking $\eta_x$ and $\eta_u$. For any time-shift $q\in \mathbb{Z}^+$, we can express $\tilde{z}(k+q)$ as
\begin{equation}
\tilde{z}(k+q)=(I_{n+m} \otimes \tau(k))\T (I_{n+m} \otimes M^q)\T \eta_z. \label{eq:ztmp}
\end{equation}
The Cayley-Hamilton theorem implies that $M^s$ can be expressed as 
$M^s=\sum_{j=0}^{s-1} c_j M^j$,
where the constants $c_j \in \mathbb{R}$, $j=0,1,\dots,s-1$ are related to the characteristic polynomial of the matrix $M$. \HChange{This fact can be combined with \eqref{eq:ztmp}, enabling us to rewrite $\tilde{z}(k+s)$ as}
\begin{equation}
\tilde{z}(k+s)=\sum_{j=0}^{s-1} c_j \tilde{z}(k+j).
\end{equation}
In other words, the property \eqref{eq:bftmp} of the basis functions implies that the trajectories $\tilde{x}(k)$, $\tilde{u}(k)$ match the trajectories of the autonomous linear time-invariant system,
\begin{align}
%\left(\begin{array}{c} \tilde{z}(k+1) \\ \vdots \\ \tilde{z}(k+s) \end{array}\right) = 
\bar{z}(k+1)=\left( \begin{array}{cccc} 0 & 1 & \dots & 0\\
0 & 0 & \dots & 0\\
\vdots & \vdots &  \ddots & \vdots \\
c_0 & c_1 & \hdots & c_{s-1} \end{array} \right)\otimes I_{n+m} %\\ 
%&
%\hspace{.6\columnwidth}\left(\begin{array}{c} \tilde{z}(k) \\ \vdots \\ \tilde{z}(k+s-1) \end{array}\right),
~\bar{z}(k),\label{eq:AutSys}
\end{align}
given the initial condition $\bar{z}(0)=(\tilde{z}(0), \dots, \tilde{z}(s-1))$, and where $\bar{z}(k):=(\tilde{z}(k), \dots, \tilde{z}(k+s-1))$. As a result, from a certain point of view, the constraint \eqref{eq:constr} describes an output admissible set of the autonomous system \eqref{eq:AutSys}. In \cite{GilbertOutput}, an algorithm is derived that computes a finite dimensional representation (or a close approximation) of the output admissible set of a given autonomous linear time-invariant system. Applied to our case this yields the following result: Under favorable circumstances (\HChange{see the next paragraph for a precise statement}), the constraint \eqref{eq:constr} is equivalent to 
\begin{equation}
g(\tilde{x}(k),\tilde{u}(k))\leq 0, \quad \forall k \in \{0,1,\dots N_\text{max}\},
\end{equation}
where $N_\text{max}$ can be computed by the recursive algorithm given in Alg.~\ref{Alg:Nmax}. For a general function $g$, the optimization that has to be carried out at each step of the algorithm is non-convex and there are therefore no guarantees that the constants $J_i$ are correctly computed. In case $g$ is affine, the optimization simplifies to a linear program that can be solved efficiently and the constants $J_i$ are guaranteed to be computed exactly (up to a given tolerance). 

As remarked in \cite{GilbertOutput} it is difficult to state general conditions on the existence of $N_\text{max}$ (that is, if the algorithm converges). According to \cite[Thm.~4.1]{GilbertOutput} an $N_\text{max} \in \mathbb{Z}^+$ is guaranteed to exist if $g$ is affine, $g(0,0)\neq 0$, and the set of all $(x,u)$ such that $g(x,u)\leq 0$ is bounded.\footnote{The observability condition, which is additionally required in \cite{GilbertOutput} is met by construction.}

%\subsection{Comments and implementation}

\begin{algorithm}
\KwData{$M, \tau(0), g$}
\KwResult{$N_\text{max} \in \mathbb{Z}^+$}
$j = (n+m)~s$\\
\While{not converged}{
	Compute $J_i$ for all $i=1,2,\dots,n_\text{c}$:
	\begin{align}
	J_i:=&\sup_{\eta_x, \eta_u} g_i(\tilde{x}(j+1), \tilde{u}(j+1))\\
	&\text{s.t. } g(\tilde{x}(k), \tilde{u}(k))\leq 0,~ \forall k \in \{0,\dots, j\}.\nonumber
	\end{align}
	\eIf{$J_i \leq 0, i=1,\dots,n_\text{c}$}{
		$N_\text{max} = j$, converged.
	}{
		$j \leftarrow j+1$
	}
}
\caption{Algorithm for computing $N_\text{max}$ (if it exists).}
\label{Alg:Nmax}
\end{algorithm}

%% file: resultingOpti.tex
\section{Resulting Optimization}\label{Sec:ResOpti}
Combining the previous results, we propose to approximate \eqref{eq:Prob} in the following way
\begin{align}
&\inf_{\eta_x \in \mathbb{R}^{ns}, \eta_u \in \mathbb{R}^{ms}} \sum_{k=0}^{\infty} l(k, \tilde{x}(k), \tilde{u}(k)) \quad \text{s.t. } \label{eq:resultingOpti}\\
&\sum_{k=0}^{\infty} (I_n \otimes \tau(k)) (\tilde{x}(k+1) - f(k, \tilde{x}(k), \tilde{u}(k))) = 0, \nonumber\\
&\tilde{x}(0)=x_0, \quad g(\tilde{x}(k), \tilde{u}(k))\leq 0, \quad \forall k\in \{0,1,\dots,N_\text{max}\}, \nonumber
\end{align}
where $\tilde{x}(k)$ is replaced by $\tilde{x}(k)=(I_n \otimes \tau(k))\T \eta_x$ and $\tilde{u}(k)$ by $\tilde{u}(k)=(I_m \otimes \tau(k))\T \eta_u$ for all $k \in \mathbb{Z}^+$.

In the special case of linear time-invariant dynamics, a quadratic cost, and affine constraints, a quadratic program of the following form is obtained
\begin{align}\label{eq:resultingOptiLin}
&\inf_{\eta_x \in \mathbb{R}^{ns}, \eta_u \in \mathbb{R}^{ms}} \eta_x\T (Q \otimes \bar{J}) \eta_x + \eta_u\T (R \otimes \bar{J}) \eta_u, \quad \text{s.t.}\\
&(I_n \otimes M\T - A \otimes I_s)\eta_x - (B \otimes I_s) \eta_u = 0,\nonumber\\
&(I_n \otimes \tau(0))\T \eta_x = x_0, \nonumber\\
& (C_\text{x} \otimes \tau(k)\T)\eta_x + (C_\text{u} \otimes \tau(k)\T)\eta_u \leq b, \nonumber \\
&\hspace{.6\columnwidth}\forall k\in \{0,\dots,N_\text{max}\} \nonumber,
\end{align}
where the matrices $Q \in \mathbb{R}^{n\times n}, R\in \mathbb{R}^{m\times m}$ encode the cost, $C_\text{x}, C_\text{u}, b$ the constraints, and $\bar{J}$ is defined as $\bar{J}:=\sum_{k=0}^{\infty} \tau(k)\tau(k)\T$.

%% file: stability.tex
\section{Closed-loop stability and recursive feasibility}\label{Sec:Stability}
Next we will discuss the question whether the resulting model predictive control algorithm, based on solving optimization problem \eqref{eq:resultingOpti} is recursively feasible and stabilizes the origin.
We restrict ourselves to the special case of linear time-invariant dynamics and a time-invariant running cost. We make the assumption that the running cost $l$ is continuous, and lower and upper bounded by two quadratic functions, i.e.
\begin{equation*}
\underline{\sigma} |x|^2 \leq l(x,u) \leq \bar{\sigma} |x|^2
\end{equation*} 
for all $x\in \mathbb{R}^n$ and $u\in \mathbb{R}^m$, where $\underline{\sigma}>0$, $\bar{\sigma} >0$. 
These assumptions ensure that the infimum in \eqref{eq:resultingOpti} is actually attained, which gives rise to (at least one) well-defined minimizer of \eqref{eq:resultingOpti}, c.f. \cite[p.~12, Ex.1.11]{Rockafellar}. The quadratic running cost as given in \eqref{eq:resultingOptiLin} satisfies these assumptions, provided that the matrix $Q$ is positive definite and $R$ is positive semi-definite. We further assume that the linear equality constraints imposing the system dynamics are regular in the sense that for any $x_0 \in \mathbb{R}^n$ there exists at least one $\eta_x\in \mathbb{R}^{ns}$ and $\eta_u\in \mathbb{R}^{ms}$ that satisfy
\begin{align*}
(I_n \otimes M\T - A \otimes I_s)\eta_x - (B \otimes I_s) \eta_u &= 0,\\
(I_n \otimes \tau(0))\T \eta_x &= x_0. \nonumber
\end{align*}
Under these assumptions we obtain the following result, which relies on the time-shift property of the basis functions. The proof can be found in 
\ifArxiv
App.~\ref{App:Proof}.
\else
\cite[App.~I]{extendedVersion}.
\fi
\begin{proposition} \label{Prop:RecFeasibilityStab}
Provided that the optimization \eqref{eq:resultingOpti} is feasible at time $0$, it remains feasible for all time steps. The resulting model predictive control algorithm asymptotically stabilizes the origin in the sense of Lyapunov.
\end{proposition}

%% file: simResults.tex
\section{\HChange{Simulation Results}}\label{Sec:Sim}
\HChange{The next section discusses the application of the proposed MPC algorithm to a quadruple integrator system. The aim is to highlight and discuss a strategy for choosing the matrix $M$ and to compare the approach to the MPC strategy presented in \cite[Ch.~3]{WangMPC} that is likewise based on the parametrization of the input.}

\HChange{The dynamics of the quadruple integrator are given by $x_\text{i}^{(4)}(t) = u_\text{i}(t)$, where $x_\text{i}(t)$ denotes the integrator state, the superscript $(4)$ denotes the fourth derivative with respect to time, and $u_\text{i}(t)$ the input. The input is constrained to $u_\text{i}(t)\in[-0.5,0.5]$ for all $t\in [0,\infty)$. A discretization with zero-order hold and a sampling time of $20$ms yields discrete-time dynamics of the form \eqref{eq:Prob} that are used as a starting point for the proposed MPC strategy. The running cost is chosen to be}
\hspace{-3pt}\begin{equation}
l(x_\text{i},\dots,\dddot{x}_\text{i},u_\text{i})=x_\text{i}^2+\dot{x}_\text{i}^2+\ddot{x}_\text{i}^2+\dddot{x}_\text{i}^2+0.05 u_\text{i}^2.
\end{equation}
\HChange{The penality on the input is deliberately chosen to be small in order generate control signals that are likely to hit the constraints. The dynamics and the running cost are motivated by the fact that the resulting optimization problem \eqref{eq:resultingOpti} reduces to a convex quadratic program. As a result, the optimization algorithm that solves \eqref{eq:resultingOpti} (respectively \eqref{eq:resultingOptiLin}) is guaranteed to return a solution close to the global minimum, irrespective of the initial guess. This enables a rigorous study with randomized initial conditions. Moreover, the approach presented in \cite{WangMPC} is based on linear time-invariant dynamics and a quadratic running cost, which is thus required for a comparison.}

\HChange{The following strategy is used for choosing the matrix $M$: We start with a parametrization given by \eqref{eq:Laguerre} and change the value of $\nu$ from 0.5s$^{-1}$ to 2s$^{-1}$ in steps of 0.1s$^{-1}$. For each value of $\nu$, we compute the averaged closed-loop cost that is achieved when running the MPC algorithm in closed-loop for 2000 steps (which amounts to 40s), starting at 100 randomized initial conditions. The initial conditions are sampled from a uniform distribution over $[-0.5,0.5]^4$ and the value $N_\text{max}$ is computed for each different choice of $M$ according to Alg.~\ref{Alg:Nmax}. For the choice $s=8$, the resulting averaged closed-loop cost as a function of $\nu$ is shown in Fig.~\ref{Fig:ChangeAlpha}. If $\nu$ is increased above 1.8s$^{-1}$ not all initial conditions are feasible. The graph reveals that the averaged closed-loop cost is a non-convex function that has multiple local minima. Due to the lack of robustness as $\nu$ increases above 1.8s$^{-1}$, the choice $\nu=0.8$s$^{-1}$ is more favorable than $\nu=1.8$s$^{-1}$, although the averaged closed-loop cost is slightly lower for $\nu=1.8$s$^{-1}$. The resulting closed-loop trajectory starting from $x_{\text{i}}(0)=\dot{x}_{\text{i}}(0)=\ddot{x}_{\text{i}}(0)=\dddot{x}_{\text{i}}(0)=0.5$ is depicted in Fig.~\ref{Fig:CLquad}, where the bang-bang behavior of the input is clearly visible.}

\newlength{\figurewidth}
\newlength{\figureheight}

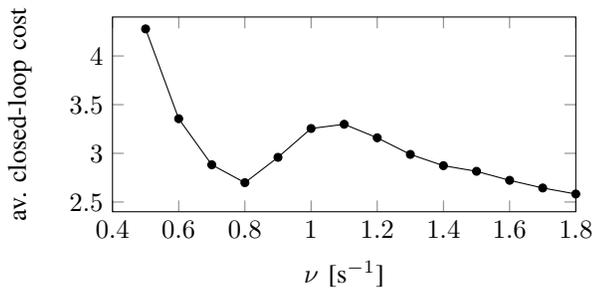
\begin{figure}
\setlength{\figurewidth}{.75\columnwidth}
\setlength{\figureheight}{.3\columnwidth}
\input{img/alphaGrid.tikz}
\caption{\HChange{The plot shows the closed-loop cost, averaged over $100$ randomized initial conditions as a function of the parameter $\nu$ describing the decay rate of the basis functions. If $\nu$ is increased above 1.8s$^{-1}$, certain initial conditions are infeasible, and cannot be stabilized by the MPC algorithm.}}
\label{Fig:ChangeAlpha}
\end{figure}

\begin{figure}
\setlength{\figurewidth}{.8\columnwidth}
\setlength{\figureheight}{.6\columnwidth}
\input{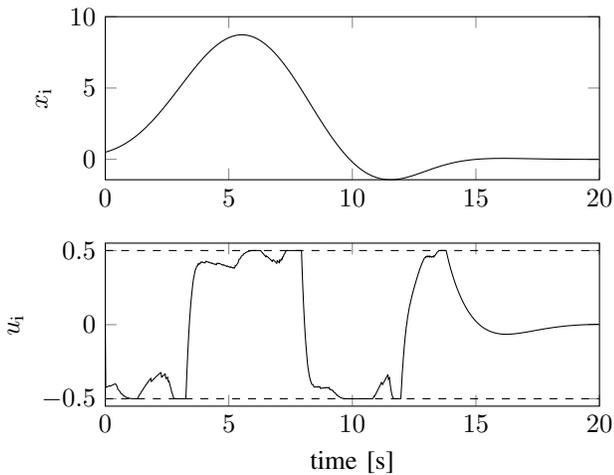}
\caption{\HChange{The plot shows the closed-loop trajectories of a quadruple integrator system that is controlled with the MPC approach presented herein, where the basis functions are chosen according to \eqref{eq:Laguerre} with $\nu=0.8$s$^{-1}$.}}
\label{Fig:CLquad}
\end{figure}

\begin{figure}
\setlength{\figurewidth}{.75\columnwidth}
\setlength{\figureheight}{.43\columnwidth}
\input{img/Benchmark.tikz}
\caption{\HChange{The plot shows a comparison between the approach presented herein (full marks) and the approach presented in \cite[Ch.~3]{WangMPC} (crosses), where the number of basis functions is varied form $s=8$ to $s=12$ (full marks) and from $s=5$ to $s=12$ (crosses). The results obtained from a dedicated MPC solver that exploits the structure of the optimization problem \eqref{eq:resultingOptiLin} are shown with empty marks. The standard deviation of the execution time (over the different randomized initial conditions) is indicated with dashed lines.}}
\label{Fig:perf}
\end{figure}
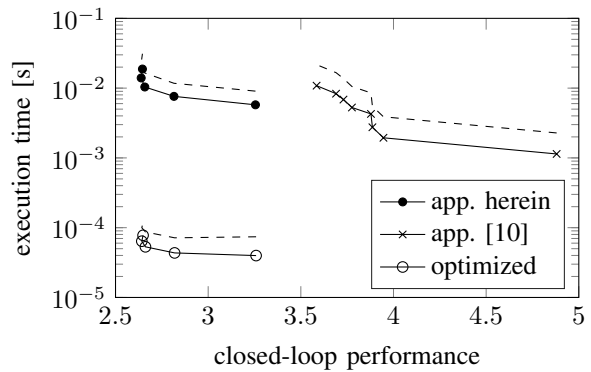

\HChange{Next, the proposed approach is compared to the approach presented in \cite[Ch.~3]{WangMPC}, where the MATLAB scripts given in \cite[p.~109,~p.~110,~p.~123]{WangMPC} are used for the implementation. The approach from \cite[Ch.~3]{WangMPC} is based on augmenting the dynamics, to parametrize the input increments (instead of the input) with Laguerre basis functions, and to eliminate the state trajectory. The value $N_\text{max}$, computed according to Alg.~\ref{Alg:Nmax}, is used as the length of the prediction horizon that is required for the approach from \cite[Ch.~3]{WangMPC}. In both cases, the resulting quadratic program is solved with the optimization routine qpOASES-v.3.2.1, \cite{qpOASES} (with the default options). Fig.~\ref{Fig:perf} compares the closed-loop performance, averaged over the 100 randomized initial conditions with the average (over the initial conditions) of the maximum (over a single simulation) execution time, when increasing the number of basis functions $s$ from 8 to 12 (for the approach presented herein) and from 5 to 12 (for the approach presented in \cite[Ch.~3]{WangMPC}). For values below $s=8$ not all trajectories are stabilized by the approach presented herein and for values below $s=5$ the performance of the approach presented in \cite[Ch.~3]{WangMPC} exceeds 5. The approach presented in \cite[Ch.~3]{WangMPC} is based on Laguerre basis functions, and therefore the same basis functions are used for the approach presented herein. The decay rate $\nu$ is set to $1\text{s}^{-1}$. It can be concluded that approach presented herein achieves a smaller cost at similar execution times. Note that in the approach from \cite[Ch.~3]{WangMPC} no terminal state-constraint and terminal cost is added, and hence closed-loop stability is not guaranteed, which contrasts the approach presented herein (see Prop.~\ref{Prop:RecFeasibilityStab}). It is observed that the approach from \cite[Ch.~3]{WangMPC} is still able to stabilize all initial conditions even for very low values of $s$ (we conducted experiments reducing $s$ up to 2). In contrast, the optimization problem \eqref{eq:resultingOptiLin} is infeasible for some initial conditions for such low values of $s$ (values $s$ below 8). This needs, however, to be put in perspective with the additional recursive feasibility and closed-loop guarantees that are inherent to \eqref{eq:resultingOptiLin}, which restrict the region where \eqref{eq:resultingOptiLin} is feasible.}

\HChange{In addition, results that are obtained with a tailored optimization routine exploiting the structure of \eqref{eq:resultingOptiLin} are also included in Fig.~\ref{Fig:perf}. The source code is available in \cite{toolbox}.}

%% file: img/alphaGrid.tikz
% Minimal pgfplots version: 1.3
%  http://www.mathworks.com/matlabcentral/fileexchange/22022-matlab2tikz
\begin{tikzpicture}

\begin{axis}[%
width=0.95092\figurewidth,
height=\figureheight,
at={(0\figurewidth,0\figureheight)},
scale only axis,
xmin=0.4,
xmax=1.8,
xlabel={$\nu$ [s$^{-1}$]},
ymin=2.4,
ymax=4.4,
ylabel={av. closed-loop cost},
legend style={legend cell align=left,align=left,draw=white!15!black}
]
\addplot [color=black,solid,mark=*,mark size=1.5pt,mark options={solid},forget plot]
  table[row sep=crcr]{%
0.5	4.27793300429324\\
0.6	3.35563583586492\\
0.7	2.88256957513767\\
0.8	2.69858332215427\\
0.9	2.95904197479563\\
1	3.25513244835296\\
1.1	3.29859642322714\\
1.2	3.15958315039339\\
1.3	2.98909609040441\\
1.4	2.87337922245474\\
1.5	2.81544015175822\\
1.6	2.72290039025211\\
1.7	2.64430180723311\\
1.8	2.58250038213853\\
};
\end{axis}
\end{tikzpicture}%

%% file: img/Benchmark.tikz
% Minimal pgfplots version: 1.3
%  http://www.mathworks.com/matlabcentral/fileexchange/22022-matlab2tikz
\begin{tikzpicture}

\begin{axis}[%
width=0.95092\figurewidth,
height=\figureheight,
at={(0\figurewidth,0\figureheight)},
scale only axis,
xmin=2.5,
xmax=5,
xlabel={closed-loop performance},
ymode=log,
ymin=1e-05,
ymax=0.1,
yminorticks=true,
ylabel={execution time [s]},
legend pos=south east,
legend style={legend cell align=left,align=left,draw=white!15!black}
]
\addplot [color=black,solid,mark=*,mark size=1.5pt,mark options={solid}]
  table[row sep=crcr]{%
3.25513244835271	0.00576507523092315\\
2.8153967131693	0.00762099987396167\\
2.65800030620747	0.0103778850768567\\
2.63844790812648	0.0139883084055248\\
2.64523462090449	0.0187752892620682\\
};
\addlegendentry{app. herein};

\addplot [color=black,dashed,forget plot]
  table[row sep=crcr]{%
3.25513244835271	0.00905594563331681\\
2.8153967131693	0.0117347304478632\\
2.65800030620747	0.0164825541748491\\
2.63844790812648	0.0229480049258688\\
2.64523462090449	0.0311097794508304\\
};

\addplot [color=black,solid,mark=x,mark size=2pt,mark options={solid}]
  table[row sep=crcr]{%
4.88037024350605	0.00113928763021708\\
3.94583826625166	0.0019420987722723\\
3.8862568140239	0.00274954534746602\\
3.87842010803843	0.00426985769458317\\
3.77571620734748	0.00527643796701825\\
3.73011785425053	0.00685314305284641\\
3.69013474995953	0.00832694640998844\\
3.58504607747838	0.0108419938920208\\
};

\addlegendentry{app. \cite{WangMPC}};

\addplot [color=black,dashed,forget plot]
  table[row sep=crcr]{%
4.88037024350605	0.00227857526043415\\
3.94583826625166	0.00388419754454461\\
3.8862568140239	0.00549909069493204\\
3.87842010803843	0.00853971538916634\\
3.77571620734748	0.0105528759340365\\
3.73011785425053	0.0137062861056928\\
3.69013474995953	0.0166538928199769\\
3.58504607747838	0.0216839877840416\\
};
\addplot [color=black,solid,mark=o,mark options={solid}]
  table[row sep=crcr]{%
3.25855295994208	3.9823185058341e-05\\
2.81882167609847	4.348527589051e-05\\
2.66142690931803	5.31701102773788e-05\\
2.64187495640607	6.39839524880664e-05\\
2.64866232581941	7.77250040746919e-05\\
};
\addlegendentry{optimized};

\addplot [color=black,dashed,forget plot]
  table[row sep=crcr]{%
3.25855295994208	7.40654381164439e-05\\
2.81882167609847	7.15911390309029e-05\\
2.66142690931803	8.68648699435311e-05\\
2.64187495640607	9.97822563906786e-05\\
2.64866232581941	0.000119408790036737\\
};
\end{axis}
\end{tikzpicture}%

%% file: expResults.tex
\section{Experimental results}\label{Sec:Exp}
The next section discusses the application of the proposed MPC algorithm to a pendulum-on-a-cart system. The aim is to swing up the pendulum from its hanging equilibrium to its upright position, while accounting for the input and state constraints. A detailed description of the experimental setup can be found in
\ifArxiv
App.~\ref{App:Exp}.
\else
\cite[App.~II]{extendedVersion}.
\fi

The cost function is chosen to be time-invariant and quadratic,
\begin{equation}
l(x,u)=20 x_\text{c}^2 + 2 \dot{x}_\text{c}^2 + 50 \varphi^2 + 2 \dot{\varphi}^2 + 10 u^2, \label{eq:cost}
\end{equation}
where relatively large penalties are given to the cart's position $x_\text{c}$, the pendulum angle $\varphi$, and the input $u$. Note that the physical units are omitted for simplicity in \eqref{eq:cost}. The cart is attached to a rail of 0.9m length, and the input (voltage applied to the motor driving the cart) is limited to $\pm24$V, hence $x_\text{c} \in [-0.45\text{m},0.45\text{m}]$ and $u\in [-24\text{V}, 24\text{V}]$.

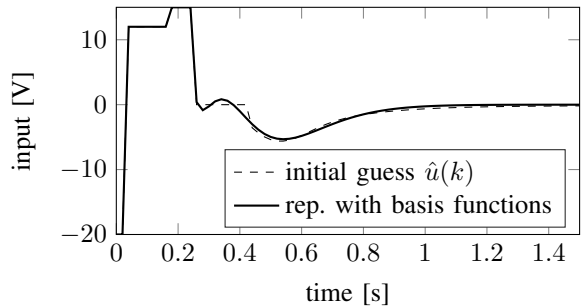
\begin{figure}
\setlength{\figurewidth}{.75\columnwidth}
\setlength{\figureheight}{.35\columnwidth}
\input{img/initialGuess.tikz}
\caption{Input trajectory represented by the basis functions that is used as an initial guess for the swing up.}
\label{Fig:InitialGuess}
\end{figure}

The basis functions are chosen to be a combination of the standard parametrization according to \eqref{eq:classicChoice} and Laguerre functions according to \eqref{eq:Laguerre}, that is,
\begin{equation}
M=\left( \begin{array}{ccc} M_\text{d} & 0 & 0 \\ 0 &\hat{\tau}(0) &\hat{M}_\text{d} \end{array}\right) \in \mathbb{R}^{19 \times 19}, 
\end{equation}
$\tau(0)=(1,0,0,\dots)\T\in \mathbb{R}^{19}$, where $M_\text{d} \in \mathbb{R}^{12 \times 12}$ is chosen according to \eqref{eq:classicChoice}, and $\hat{M}_\text{d}\in \mathbb{R}^{7\times 7}$ and $\hat{\tau}(0)$ are chosen according to \eqref{eq:Laguerre} with $\nu=14$s$^{-1}$. 
\HChange{The choice is motivated by the fact that a typical swing-up trajectory (see Fig.~\ref{Fig:InitialGuess}) involves a bang-bang control input for a bit more than 0.2s (hence roughly 12 time steps) followed by smaller inputs that guide the pendulum to its upright position. These smaller inputs are well approximated by Laguerre basis functions with a relatively small time constant of $1/(14)$s. A small time constant is desirable due to the fact that the resulting value of $N_\text{max}$, and hence the number of inequalities in the corresponding optimization problem, is relatively small.}
The basis functions are chosen to be orthogonal, that is, they are normalized such that $\sum_{k=0}^{\infty} \tau(k)\tau(k)\T = I_s$. Running algorithm Alg.~\ref{Alg:Nmax} subject to the input (limited voltage) and state constraint (limited rail length) results in $N_\text{max}=30$. 

A rough initial guess for the swing-up trajectories (state and input) is generated via simulation, where the input $\hat{u}(k)$ is obtained by combining step inputs with a linear quadratic regulator for catching the pendulum in upright position. The input trajectory is shown in Fig.~\ref{Fig:InitialGuess} (dashed line). The trajectories are then represented using the basis functions. In case of the input trajectory, the resulting parameter vector is given by 
$\hat{\eta}_u=\sum_{k=0}^{\infty} \tau(k) \hat{u}(k)$, and the resulting trajectory $\tau(k)\T \hat{\eta}_u$ is shown in Fig.~\ref{Fig:InitialGuess} (solid line). The representation of the state trajectories are obtained analogously.
The state constraint (limited rail length) is not considered for the generation of the initial guess.

The dynamics are imposed using the variational formulation given by \eqref{eq:dynTmp}, where the infinite sum is truncated after $k=150$. The truncation is motivated by the fact that the contributions for $k>42$ are negligible, as the basis function decay with a time constant of $1/\nu=0.071$s (that is, 4 samples result in a decay of roughly 67\%).

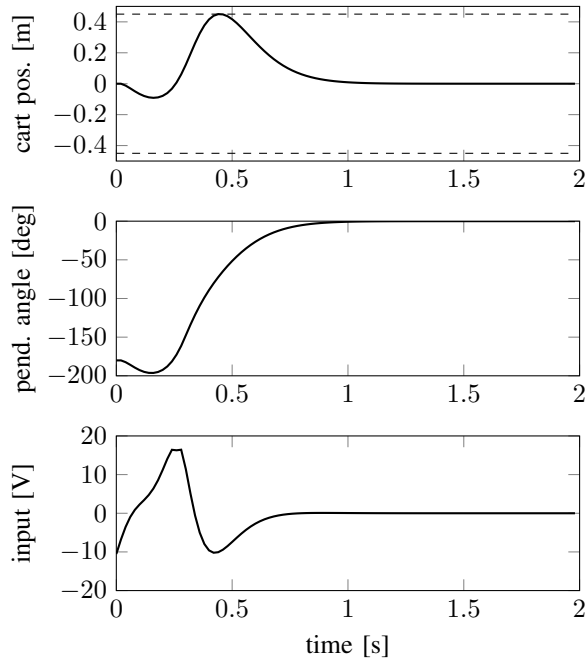
\begin{figure}
\setlength{\figurewidth}{.75\columnwidth}
\setlength{\figureheight}{0.9\columnwidth}
\input{img/swingUpNom.tikz}
\caption{Nominal trajectories resulting from optimizing the optimal control problem at time $t=0$.}
\label{Fig:SwingUpNom}
\end{figure}

The nonlinear optimization problem is solved with Ipopt, \cite{Ipopt} (version 3.12.10) and interfaced through C++. The functions for evaluating the derivatives of the Lagrangian and the equality constraints are generated with Casadi, \cite{Casadi} (version 3.4.4). Ipopt is run with the default settings, a relative tolerance of 1e-6, and with a maximum number of 8 iterations (during warmstarts). The time-shift property of the basis functions is exploited for warm-starting the optimization during the swing up. The nominal open-loop swing-up trajectory (as obtained by solving the optimization at the first time instant) is shown in Fig.~\ref{Fig:SwingUpNom}.

\begin{figure}
\setlength{\figurewidth}{.75\columnwidth}
\setlength{\figureheight}{1.15\columnwidth}
\input{img/swingUp.tikz}
\caption{Trajectories resulting from a successful swing up.}
\label{Fig:SuccSwingUp}
\end{figure}
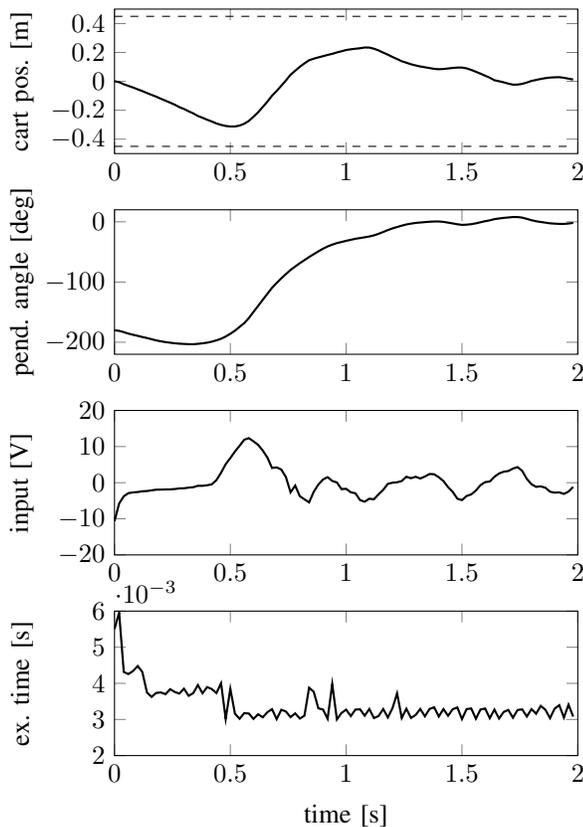

The resulting swing-up trajectories are shown in Fig.~\ref{Fig:SuccSwingUp}. After roughly 1s the pendulum reaches its upright position. During the whole swing-up trajectory the input and state constraints are accounted for, and the execution time of the optimization remains below 0.02s. 

Note the difference between the planned open-loop trajectory and resulting closed-loop trajectory. For instance, the constraint at $x_\text{c}=0.45$m is reached during the open-loop trajectory, whereas in closed-loop $x_\text{c}=0.45$m is never reached. This discrepancy is most likely due to modeling errors. In particular, the dry friction between the cart and the rail, which is substantial, is difficult to model accurately and to compensate for. Nevertheless, the MPC controller is able to perform the swing up. 
\HChange{The robustness of the optimization with respect to the initial guess is further discussed and quantified in 
\ifArxiv
App.~\ref{App:Exp}.
\else
\cite[App.~II]{extendedVersion}.
\fi
}

The above choice of basis functions enables a reduction of the degrees of freedom. In case a standard parametrization according to \eqref{eq:classicChoice} is used, 250 optimization variables are needed to capture a prediction horizon of 1s, which is roughly required for performing a swing up. The parametrization allows to capture about the same time horizon with only 85 optimization variables.

%% file: img/initialGuess.tikz
% Minimal pgfplots version: 1.3
%  http://www.mathworks.com/matlabcentral/fileexchange/22022-matlab2tikz
\begin{tikzpicture}

\begin{axis}[%
width=0.95092\figurewidth,
height=\figureheight,
at={(0\figurewidth,0\figureheight)},
scale only axis,
xmin=0,
xmax=1.5,
xlabel={time [s]},
ymin=-20,
ymax=15,
ylabel={input [V]},
legend style={legend cell align=left,align=left,draw=white!15!black},
legend pos={south east}
]
\addplot [color=black,dashed]
  table[row sep=crcr]{%
0	-20\\
0.02	-20\\
0.04	12\\
0.06	12\\
0.08	12\\
0.1	12\\
0.12	12\\
0.14	12\\
0.16	12\\
0.18	15\\
0.2	15\\
0.22	15\\
0.24	15\\
0.26	0\\
0.28	0\\
0.3	0\\
0.32	0\\
0.34	0\\
0.36	0\\
0.38	0\\
0.4	0\\
0.42	0\\
0.44	-3.56196590103068\\
0.46	-4.35247176124713\\
0.48	-4.96947771650957\\
0.5	-5.38871746952111\\
0.52	-5.60066354267578\\
0.54	-5.61237779749226\\
0.56	-5.44728114642287\\
0.58	-5.14231369212136\\
0.6	-4.7425443182477\\
0.62	-4.29419980214973\\
0.64	-3.83787273943219\\
0.66	-3.40377483282221\\
0.68	-3.0101081992901\\
0.7	-2.66434961342217\\
0.72	-2.3662552213021\\
0.74	-2.11118271869866\\
0.76	-1.89276979995191\\
0.78	-1.70463473579128\\
0.8	-1.5412105899609\\
0.82	-1.39799762946045\\
0.84	-1.27150544627393\\
0.86	-1.15907226402219\\
0.88	-1.05866521801824\\
0.9	-0.968707254464704\\
0.92	-0.887943652964999\\
0.94	-0.815346060710347\\
0.96	-0.750046648283808\\
0.98	-0.691294394279294\\
1	-0.638426707772549\\
1.02	-0.590851223618123\\
1.04	-0.548034092254825\\
1.06	-0.509492263792889\\
1.08	-0.474788132185292\\
1.1	-0.443525512182541\\
1.12	-0.415346331883452\\
1.14	-0.389927691735241\\
1.16	-0.366979109726982\\
1.18	-0.346239874107417\\
1.2	-0.327476482110936\\
1.22	-0.310480171943467\\
1.24	-0.295064566693681\\
1.26	-0.281063450301079\\
1.28	-0.268328692059361\\
1.3	-0.256728330383239\\
1.32	-0.246144820503349\\
1.34	-0.236473445334255\\
1.36	-0.227620884411257\\
1.38	-0.219503932615962\\
1.4	-0.212048358330656\\
1.42	-0.205187889511503\\
1.44	-0.198863315755842\\
1.46	-0.193021694569111\\
1.48	-0.187615650544821\\
1.5	-0.182602756918693\\
};
\addlegendentry{initial guess $\hat{u}(k)$};

\addplot [color=black,thick,solid]
  table[row sep=crcr]{%
0	-20\\
0.02	-20\\
0.04	12\\
0.06	12\\
0.08	12\\
0.1	12\\
0.12	12\\
0.14	12\\
0.16	12\\
0.18	15\\
0.2	15\\
0.22	15\\
0.24	15\\
0.26	0.393402227146159\\
0.28	-0.848385923088417\\
0.3	-0.255933247446653\\
0.32	0.513715984380408\\
0.34	0.836793377390331\\
0.36	0.603958839099092\\
0.38	-0.0714424525487396\\
0.4	-1.00850587291184\\
0.42	-2.03408513408943\\
0.44	-3.01333711933582\\
0.46	-3.85661778224341\\
0.48	-4.51509195999266\\
0.5	-4.97199868048245\\
0.52	-5.23323326036545\\
0.54	-5.31898057888812\\
0.56	-5.25704787518321\\
0.58	-5.07797051779258\\
0.6	-4.81169114128287\\
0.62	-4.48551068829621\\
0.64	-4.12299938704608\\
0.66	-3.74358907125419\\
0.68	-3.36261836923511\\
0.7	-2.99165450355952\\
0.72	-2.63896258732444\\
0.74	-2.31003253299208\\
0.76	-2.00810452414384\\
0.78	-1.73465714071202\\
0.8	-1.48983886853499\\
0.82	-1.27283516867808\\
0.84	-1.08217074889896\\
0.86	-0.915951214608466\\
0.88	-0.772050729040755\\
0.9	-0.648253343748011\\
0.92	-0.542355768599913\\
0.94	-0.452238897459934\\
0.96	-0.375914646752993\\
0.98	-0.311553772365682\\
1	-0.257499417977983\\
1.02	-0.21227028333267\\
1.04	-0.174556521678032\\
1.06	-0.143210798455345\\
1.08	-0.117236371565821\\
1.1	-0.095773582350499\\
1.12	-0.0780857663032389\\
1.14	-0.0635452919757572\\
1.16	-0.051620203497396\\
1.18	-0.0418617650799349\\
1.2	-0.0338930743004458\\
1.22	-0.0273988156330944\\
1.24	-0.0221161587716123\\
1.26	-0.0178267611967009\\
1.28	-0.0143498058452422\\
1.3	-0.0115359883362311\\
1.32	-0.00926236061218666\\
1.34	-0.00742793643227303\\
1.36	-0.00594996689396898\\
1.38	-0.00476079955498792\\
1.4	-0.0038052416657746\\
1.42	-0.00303835570647951\\
1.44	-0.00242362329223302\\
1.46	-0.00193142119048515\\
1.48	-0.00153776044384735\\
1.5	-0.00122324627060296\\
};
\addlegendentry{rep. with basis functions};

\end{axis}
\end{tikzpicture}%

%% file: img/swingUpNom.tikz
% Minimal pgfplots version: 1.3
%  http://www.mathworks.com/matlabcentral/fileexchange/22022-matlab2tikz
\begin{tikzpicture}

\begin{axis}[%
width=0.95092\figurewidth,
height=0.264706\figureheight,
at={(0\figurewidth,0.735294\figureheight)},
scale only axis,
xmin=0,
xmax=2,
xlabel={},
xlabel near ticks,
ymin=-0.5,
ymax=0.5,
ylabel={cart pos. [m]}
]
\addplot [color=black,thick,solid,forget plot]
  table[row sep=crcr]{%
0	8.2431563755831e-33\\
0.02	-9.32097530652007e-38\\
0.04	-0.0119883305367444\\
0.06	-0.0298631832471675\\
0.08	-0.048969933479104\\
0.1	-0.0661330478502521\\
0.12	-0.0794935128339076\\
0.14	-0.0880799174841691\\
0.16	-0.0912975451359523\\
0.18	-0.0885089593289387\\
0.2	-0.0787614074555066\\
0.22	-0.0606632837858837\\
0.24	-0.0324539075109175\\
0.26	0.00758811077856617\\
0.28	0.0601951360768159\\
0.3	0.122892926717955\\
0.32	0.192399455110647\\
0.34	0.261935845993066\\
0.36	0.325041761231503\\
0.38	0.377070336465726\\
0.4	0.415479507890175\\
0.42	0.439566085974431\\
0.44	0.450000009999603\\
0.46	0.448343041396419\\
0.48	0.436635670192203\\
0.5	0.417080335313579\\
0.52	0.39182035622779\\
0.54	0.362800991366541\\
0.56	0.331694812054093\\
0.58	0.299873894613009\\
0.6	0.268413715855139\\
0.62	0.238116700790777\\
0.64	0.20954638883007\\
0.66	0.18306581888362\\
0.68	0.158875868594075\\
0.7	0.137050919996539\\
0.72	0.117570420590408\\
0.74	0.100345742944463\\
0.76	0.0852422967334696\\
0.78	0.0720971857129606\\
0.8	0.060732888119343\\
0.82	0.0509675196707866\\
0.84	0.0426222498893436\\
0.86	0.0355264115055609\\
0.88	0.0295207881995264\\
0.9	0.0244595008638294\\
0.92	0.0202108454450578\\
0.94	0.0166573714628713\\
0.96	0.0136954324426192\\
0.98	0.0112343890781034\\
1	0.00919560328613793\\
1.02	0.00751132611055633\\
1.04	0.00612355400952611\\
1.06	0.00498290557655937\\
1.08	0.00404755331633927\\
1.1	0.00328223186495262\\
1.12	0.00265733422667568\\
1.14	0.00214810050612516\\
1.16	0.00173389865540808\\
1.18	0.00139759343831069\\
1.2	0.00112499773456124\\
1.22	0.000904399143020101\\
1.24	0.000726154337572656\\
1.26	0.0005823435848761\\
1.28	0.000466478097217843\\
1.3	0.000373253352770459\\
1.32	0.000298342085806196\\
1.34	0.000238221270889623\\
1.36	0.000190028055817219\\
1.38	0.00015144021003892\\
1.4	0.000120577230696584\\
1.42	9.59187770000849e-05\\
1.44	7.62375805736443e-05\\
1.46	6.05444035289115e-05\\
1.48	4.80429887493627e-05\\
1.5	3.8093271182206e-05\\
1.52	3.01813987259102e-05\\
1.54	2.38953509115318e-05\\
1.56	1.8905147452575e-05\\
1.58	1.49468112201152e-05\\
1.6	1.18093953776589e-05\\
1.62	9.32450604518646e-06\\
1.64	7.35785334805375e-06\\
1.66	5.8024480606687e-06\\
1.68	4.57313092580237e-06\\
1.7	3.60217941990961e-06\\
1.72	2.8357842260785e-06\\
1.74	2.23122666502029e-06\\
1.76	1.75462026152955e-06\\
1.78	1.37910570597243e-06\\
1.8	1.08340973019957e-06\\
1.82	8.50695710213139e-07\\
1.84	6.67647846493196e-07\\
1.86	5.23742148425954e-07\\
1.88	4.10666650803205e-07\\
1.9	3.21860721025673e-07\\
1.92	2.52149306823364e-07\\
1.94	1.97452797653669e-07\\
1.96	1.54557050580569e-07\\
1.98	1.20931244582582e-07\\
};
\addplot [color=black,dashed,forget plot]
  table[row sep=crcr]{%
0	-0.45\\
0.02	-0.45\\
0.04	-0.45\\
0.06	-0.45\\
0.08	-0.45\\
0.1	-0.45\\
0.12	-0.45\\
0.14	-0.45\\
0.16	-0.45\\
0.18	-0.45\\
0.2	-0.45\\
0.22	-0.45\\
0.24	-0.45\\
0.26	-0.45\\
0.28	-0.45\\
0.3	-0.45\\
0.32	-0.45\\
0.34	-0.45\\
0.36	-0.45\\
0.38	-0.45\\
0.4	-0.45\\
0.42	-0.45\\
0.44	-0.45\\
0.46	-0.45\\
0.48	-0.45\\
0.5	-0.45\\
0.52	-0.45\\
0.54	-0.45\\
0.56	-0.45\\
0.58	-0.45\\
0.6	-0.45\\
0.62	-0.45\\
0.64	-0.45\\
0.66	-0.45\\
0.68	-0.45\\
0.7	-0.45\\
0.72	-0.45\\
0.74	-0.45\\
0.76	-0.45\\
0.78	-0.45\\
0.8	-0.45\\
0.82	-0.45\\
0.84	-0.45\\
0.86	-0.45\\
0.88	-0.45\\
0.9	-0.45\\
0.92	-0.45\\
0.94	-0.45\\
0.96	-0.45\\
0.98	-0.45\\
1	-0.45\\
1.02	-0.45\\
1.04	-0.45\\
1.06	-0.45\\
1.08	-0.45\\
1.1	-0.45\\
1.12	-0.45\\
1.14	-0.45\\
1.16	-0.45\\
1.18	-0.45\\
1.2	-0.45\\
1.22	-0.45\\
1.24	-0.45\\
1.26	-0.45\\
1.28	-0.45\\
1.3	-0.45\\
1.32	-0.45\\
1.34	-0.45\\
1.36	-0.45\\
1.38	-0.45\\
1.4	-0.45\\
1.42	-0.45\\
1.44	-0.45\\
1.46	-0.45\\
1.48	-0.45\\
1.5	-0.45\\
1.52	-0.45\\
1.54	-0.45\\
1.56	-0.45\\
1.58	-0.45\\
1.6	-0.45\\
1.62	-0.45\\
1.64	-0.45\\
1.66	-0.45\\
1.68	-0.45\\
1.7	-0.45\\
1.72	-0.45\\
1.74	-0.45\\
1.76	-0.45\\
1.78	-0.45\\
1.8	-0.45\\
1.82	-0.45\\
1.84	-0.45\\
1.86	-0.45\\
1.88	-0.45\\
1.9	-0.45\\
1.92	-0.45\\
1.94	-0.45\\
1.96	-0.45\\
1.98	-0.45\\
};
\addplot [color=black,dashed,forget plot]
  table[row sep=crcr]{%
0	0.45\\
0.02	0.45\\
0.04	0.45\\
0.06	0.45\\
0.08	0.45\\
0.1	0.45\\
0.12	0.45\\
0.14	0.45\\
0.16	0.45\\
0.18	0.45\\
0.2	0.45\\
0.22	0.45\\
0.24	0.45\\
0.26	0.45\\
0.28	0.45\\
0.3	0.45\\
0.32	0.45\\
0.34	0.45\\
0.36	0.45\\
0.38	0.45\\
0.4	0.45\\
0.42	0.45\\
0.44	0.45\\
0.46	0.45\\
0.48	0.45\\
0.5	0.45\\
0.52	0.45\\
0.54	0.45\\
0.56	0.45\\
0.58	0.45\\
0.6	0.45\\
0.62	0.45\\
0.64	0.45\\
0.66	0.45\\
0.68	0.45\\
0.7	0.45\\
0.72	0.45\\
0.74	0.45\\
0.76	0.45\\
0.78	0.45\\
0.8	0.45\\
0.82	0.45\\
0.84	0.45\\
0.86	0.45\\
0.88	0.45\\
0.9	0.45\\
0.92	0.45\\
0.94	0.45\\
0.96	0.45\\
0.98	0.45\\
1	0.45\\
1.02	0.45\\
1.04	0.45\\
1.06	0.45\\
1.08	0.45\\
1.1	0.45\\
1.12	0.45\\
1.14	0.45\\
1.16	0.45\\
1.18	0.45\\
1.2	0.45\\
1.22	0.45\\
1.24	0.45\\
1.26	0.45\\
1.28	0.45\\
1.3	0.45\\
1.32	0.45\\
1.34	0.45\\
1.36	0.45\\
1.38	0.45\\
1.4	0.45\\
1.42	0.45\\
1.44	0.45\\
1.46	0.45\\
1.48	0.45\\
1.5	0.45\\
1.52	0.45\\
1.54	0.45\\
1.56	0.45\\
1.58	0.45\\
1.6	0.45\\
1.62	0.45\\
1.64	0.45\\
1.66	0.45\\
1.68	0.45\\
1.7	0.45\\
1.72	0.45\\
1.74	0.45\\
1.76	0.45\\
1.78	0.45\\
1.8	0.45\\
1.82	0.45\\
1.84	0.45\\
1.86	0.45\\
1.88	0.45\\
1.9	0.45\\
1.92	0.45\\
1.94	0.45\\
1.96	0.45\\
1.98	0.45\\
};
\end{axis}

\begin{axis}[%
width=0.95092\figurewidth,
height=0.264706\figureheight,
at={(0\figurewidth,0\figureheight)},
scale only axis,
xmin=0,
xmax=2,
xlabel={time [s]},
xlabel near ticks,
ymin=-20,
ymax=20,
ylabel={input [V]}
]
\addplot [color=black,thick,solid,forget plot]
  table[row sep=crcr]{%
0	-10.53692870498\\
0.02	-6.79884608724063\\
0.04	-3.54582342887528\\
0.06	-0.978739051689967\\
0.08	0.88124155275057\\
0.1	2.24334043789249\\
0.12	3.42428301188735\\
0.14	4.72686054062159\\
0.16	6.38546314366597\\
0.18	8.54207605949247\\
0.2	11.1915365824962\\
0.22	14.0550221374374\\
0.24	16.4158657810087\\
0.26	16.2875828436138\\
0.28	16.4353859820059\\
0.3	11.4259907722978\\
0.32	5.12692826708447\\
0.34	-0.593373688964754\\
0.36	-4.99678865681799\\
0.38	-7.9469983084218\\
0.4	-9.59705839949656\\
0.42	-10.2118771722194\\
0.44	-10.0731825025061\\
0.46	-9.43354888650926\\
0.48	-8.49850572047345\\
0.5	-7.4238244079461\\
0.52	-6.32025555979554\\
0.54	-5.26125185285509\\
0.56	-4.29123265371144\\
0.58	-3.43316644922024\\
0.6	-2.69495909396182\\
0.62	-2.07453131379013\\
0.64	-1.56367161569222\\
0.66	-1.15083975875184\\
0.68	-0.823120813832839\\
0.7	-0.56752102169908\\
0.72	-0.371772312423517\\
0.74	-0.224782877307843\\
0.76	-0.116842188093081\\
0.78	-0.0396630387768609\\
0.8	0.0136784411572796\\
0.82	0.0488611995570308\\
0.84	0.0704659436524764\\
0.86	0.0821300539235493\\
0.88	0.0866981006924697\\
0.9	0.0863606078967353\\
0.92	0.0827773484497738\\
0.94	0.0771838771304227\\
0.96	0.0704815419223306\\
0.98	0.0633121209323913\\
1	0.0561187048040775\\
1.02	0.049194628604643\\
1.04	0.0427222572974753\\
1.06	0.0368033192305422\\
1.08	0.0314823136270532\\
1.1	0.0267643249078847\\
1.12	0.0226283805963746\\
1.14	0.0190373036885654\\
1.16	0.0159448418910148\\
1.18	0.0133007082286228\\
1.2	0.0110540408440929\\
1.22	0.00915568343199877\\
1.24	0.00755959989340341\\
1.26	0.00622366527891324\\
1.28	0.00511001762191646\\
1.3	0.00418510964048468\\
1.32	0.00341956347773273\\
1.34	0.00278790385846053\\
1.36	0.00226822371234862\\
1.38	0.00184182013807093\\
1.4	0.00149282646698566\\
1.42	0.00120785723525968\\
1.44	0.000975676365913112\\
1.46	0.000786894218847039\\
1.48	0.000633695929834842\\
1.5	0.000509601270457843\\
1.52	0.000409254842392697\\
1.54	0.000328244558303395\\
1.56	0.000262945896153676\\
1.58	0.000210389222646437\\
1.6	0.00016814747469302\\
1.62	0.000134241599467548\\
1.64	0.000107061336306814\\
1.66	8.52991440043266e-05\\
1.68	6.78953118560783e-05\\
1.7	5.39925267646471e-05\\
1.72	4.28983920075161e-05\\
1.74	3.40546001389225e-05\\
1.76	2.70116499467269e-05\\
1.78	2.14081643752949e-05\\
1.8	1.69540130732456e-05\\
1.82	1.34165707529857e-05\\
1.84	1.06095523439027e-05\\
1.86	8.38395970000429e-06\\
1.88	6.62075417384312e-06\\
1.9	5.22493644665736e-06\\
1.92	4.12077126658566e-06\\
1.94	3.24794171448027e-06\\
1.96	2.5584566598605e-06\\
1.98	2.01416740096702e-06\\
};
\end{axis}

\begin{axis}[%
width=0.95092\figurewidth,
height=0.264706\figureheight,
at={(0\figurewidth,0.367647\figureheight)},
scale only axis,
xmin=0,
xmax=2,
xlabel={},
ymin=-200,
ymax=0,
ylabel={pend. angle [deg]},
xlabel near ticks
]
\addplot [color=black,thick,solid,forget plot]
  table[row sep=crcr]{%
0	-180.057295779513\\
0.02	-180.057295779513\\
0.04	-182.351585277103\\
0.06	-185.772036772411\\
0.08	-189.397300544603\\
0.1	-192.576812203255\\
0.12	-194.915458497553\\
0.14	-196.198515958752\\
0.16	-196.295050128554\\
0.18	-195.077841968827\\
0.2	-192.371150604896\\
0.22	-187.925451347005\\
0.24	-181.428061968606\\
0.26	-172.58349240524\\
0.28	-161.31550787671\\
0.3	-147.664514359073\\
0.32	-133.944425499301\\
0.34	-121.104125053002\\
0.36	-109.411803933683\\
0.38	-98.8329314997989\\
0.4	-89.227547619403\\
0.42	-80.4444009780642\\
0.44	-72.3587831726514\\
0.46	-64.881638202803\\
0.48	-57.9556242077664\\
0.5	-51.5466050987173\\
0.52	-45.634812197573\\
0.54	-40.2075120170658\\
0.56	-35.2537236506604\\
0.58	-30.760891766504\\
0.6	-26.7131556603297\\
0.62	-23.090788986587\\
0.64	-19.8704167808981\\
0.66	-17.025688194711\\
0.68	-14.5281637489874\\
0.7	-12.348249556983\\
0.72	-10.4560716581842\\
0.74	-8.82223007598071\\
0.76	-7.41840563322902\\
0.78	-6.21781510877373\\
0.8	-5.19552438676811\\
0.82	-4.32863705303226\\
0.84	-3.59637929669685\\
0.86	-2.98010246268884\\
0.88	-2.46322329054334\\
0.9	-2.03111957679964\\
0.92	-1.6709962711852\\
0.94	-1.37173422785828\\
0.96	-1.1237312067428\\
0.98	-0.918742379577293\\
1	-0.749725594434886\\
1.02	-0.610694999652102\\
1.04	-0.496585304245216\\
1.06	-0.403127922773626\\
1.08	-0.326739477328044\\
1.1	-0.264422565779826\\
1.12	-0.213678313675549\\
1.14	-0.172429971232595\\
1.16	-0.138956665631914\\
1.18	-0.111836346035427\\
1.2	-0.0898969430439025\\
1.22	-0.0721747885126263\\
1.24	-0.057879392347043\\
1.26	-0.0463637398550265\\
1.28	-0.0370993487849638\\
1.3	-0.0296554037771004\\
1.32	-0.0236813637068628\\
1.34	-0.0188925116966145\\
1.36	-0.0150579867865355\\
1.38	-0.0119908994811797\\
1.4	-0.00954019024558687\\
1.42	-0.00758394050126554\\
1.44	-0.00602389000284379\\
1.46	-0.00478095305186648\\
1.48	-0.00379155930537001\\
1.5	-0.00300467348432586\\
1.52	-0.00237937260834532\\
1.54	-0.00188287998956054\\
1.56	-0.00148897259107604\\
1.58	-0.00117669293552581\\
1.6	-0.000929308935761029\\
1.62	-0.000733475167281867\\
1.64	-0.000578557522667214\\
1.66	-0.000456090153559677\\
1.68	-0.000359339350217705\\
1.7	-0.000282953733202466\\
1.72	-0.000222684007526034\\
1.74	-0.00017515870140244\\
1.76	-0.000137704901618743\\
1.78	-0.000108205107766729\\
1.8	-8.49830435450118e-05\\
1.82	-6.67126560826995e-05\\
1.84	-5.23456626162595e-05\\
1.86	-4.10539165035436e-05\\
1.88	-3.2183601541278e-05\\
1.9	-2.52188577694587e-05\\
1.92	-1.97529203572994e-05\\
1.94	-1.54652378059179e-05\\
1.96	-1.21033445430256e-05\\
1.98	-9.46851064868068e-06\\
};
\end{axis}
\end{tikzpicture}%

%% file: img/swingUp.tikz
% Minimal pgfplots version: 1.3
%  http://www.mathworks.com/matlabcentral/fileexchange/22022-matlab2tikz
\begin{tikzpicture}

\begin{axis}[%
width=0.95092\figurewidth,
height=0.193548\figureheight,
at={(0\figurewidth,0.537634\figureheight)},
scale only axis,
xmin=0,
xmax=2,
xlabel={},
ymin=-220,
ymax=20,
ylabel={pend. angle [deg]}
]
\addplot [color=black,thick,solid,forget plot]
  table[row sep=crcr]{%
0	-180.042819795139\\
0.02	-181.122272281166\\
0.04	-183.184576568959\\
0.06	-185.739968335243\\
0.08	-187.617035367871\\
0.1	-189.36455664302\\
0.12	-191.349568924251\\
0.14	-193.029423883795\\
0.16	-194.974844781382\\
0.18	-196.969196274673\\
0.2	-198.464616119965\\
0.22	-199.778809414656\\
0.24	-200.765958399887\\
0.26	-201.744971384427\\
0.28	-202.434010428852\\
0.3	-202.824423870454\\
0.32	-203.230364468304\\
0.34	-203.359910225783\\
0.36	-202.667777209265\\
0.38	-201.819627785133\\
0.4	-200.728143185409\\
0.42	-199.074185918205\\
0.44	-196.902790466218\\
0.46	-194.243177677\\
0.48	-190.519467670665\\
0.5	-185.990522779054\\
0.52	-180.765949828374\\
0.54	-174.534690031649\\
0.56	-168.28985113518\\
0.58	-159.887883435822\\
0.6	-149.610682168781\\
0.62	-139.674505381402\\
0.64	-129.491956742116\\
0.66	-119.571135223647\\
0.68	-109.632036351512\\
0.7	-101.558086989867\\
0.72	-93.960723922212\\
0.74	-86.725642068418\\
0.76	-79.9233725330659\\
0.78	-74.2616900804806\\
0.8	-68.701249270295\\
0.82	-63.6389379671961\\
0.84	-58.4879900931905\\
0.86	-53.724877350709\\
0.88	-49.0663294058388\\
0.9	-44.694260358534\\
0.92	-40.8466195811125\\
0.94	-37.9044240073362\\
0.96	-35.2959190534441\\
0.98	-33.5027903378027\\
1	-31.5355970440005\\
1.02	-29.8636489020293\\
1.04	-28.14792678451\\
1.06	-26.9560026832994\\
1.08	-25.717210634447\\
1.1	-24.3196519805639\\
1.12	-22.5273254058357\\
1.14	-19.8887974634787\\
1.16	-16.8852381098439\\
1.18	-13.6964987968228\\
1.2	-10.9201745047369\\
1.22	-8.60015380069317\\
1.24	-6.39632893750197\\
1.26	-4.36914696254961\\
1.28	-2.99593901495956\\
1.3	-1.80636404070895\\
1.32	-1.15067113996123\\
1.34	-0.476815477107871\\
1.36	0.144270772813941\\
1.38	0.518584100372908\\
1.4	0.57513503475232\\
1.42	-0.0868031059623197\\
1.44	-1.34501842406961\\
1.46	-2.6676341983496\\
1.48	-4.05109809047249\\
1.5	-5.04380476631615\\
1.52	-4.72684451404978\\
1.54	-3.90946292351615\\
1.56	-2.51706088978922\\
1.58	-0.717343159503791\\
1.6	1.04364262383079\\
1.62	2.48445959124628\\
1.64	4.4413969405156\\
1.66	5.56605579657789\\
1.68	6.51928568033705\\
1.7	7.23009712097635\\
1.72	7.95208123862069\\
1.74	7.94520574507913\\
1.76	6.96596357742104\\
1.78	4.89437737334603\\
1.8	2.68419267862888\\
1.82	0.964058786087123\\
1.84	-0.283556812810244\\
1.86	-1.34616433965987\\
1.88	-2.47139615351729\\
1.9	-3.04280696260126\\
1.92	-3.42892322073992\\
1.94	-3.46140992772384\\
1.96	-2.86066367952917\\
1.98	-1.79530595526292\\
};
\end{axis}

\begin{axis}[%
width=0.95092\figurewidth,
height=0.193548\figureheight,
at={(0\figurewidth,0\figureheight)},
scale only axis,
xmin=0,
xmax=2,
xlabel={time [s]},
ymin=0.002,
ymax=0.006,
ylabel={ex. time [s]}
]
\addplot [color=black,thick,solid,forget plot]
  table[row sep=crcr]{%
0	0.005504\\
0.02	0.005959\\
0.04	0.004314\\
0.06	0.004257\\
0.08	0.004341\\
0.1	0.00448\\
0.12	0.004316\\
0.14	0.003749\\
0.16	0.003625\\
0.18	0.003735\\
0.2	0.003748\\
0.22	0.003699\\
0.24	0.003839\\
0.26	0.003762\\
0.28	0.003716\\
0.3	0.003852\\
0.32	0.003655\\
0.34	0.00373\\
0.36	0.003952\\
0.38	0.003725\\
0.4	0.003897\\
0.42	0.003841\\
0.44	0.00373\\
0.46	0.004002\\
0.48	0.003018\\
0.5	0.003847\\
0.52	0.003159\\
0.54	0.003019\\
0.56	0.003175\\
0.58	0.003165\\
0.6	0.003015\\
0.62	0.003145\\
0.64	0.003059\\
0.66	0.003201\\
0.68	0.003286\\
0.7	0.003008\\
0.72	0.003268\\
0.74	0.00322\\
0.76	0.003063\\
0.78	0.003169\\
0.8	0.003024\\
0.82	0.003105\\
0.84	0.003878\\
0.86	0.003768\\
0.88	0.003307\\
0.9	0.003276\\
0.92	0.003019\\
0.94	0.003979\\
0.96	0.003002\\
0.98	0.00318\\
1	0.003272\\
1.02	0.003093\\
1.04	0.003274\\
1.06	0.003281\\
1.08	0.003103\\
1.1	0.003289\\
1.12	0.003038\\
1.14	0.003096\\
1.16	0.003289\\
1.18	0.003056\\
1.2	0.003311\\
1.22	0.003721\\
1.24	0.003042\\
1.26	0.003297\\
1.28	0.003041\\
1.3	0.003125\\
1.32	0.003291\\
1.34	0.003087\\
1.36	0.00329\\
1.38	0.003338\\
1.4	0.003081\\
1.42	0.003288\\
1.44	0.003102\\
1.46	0.003283\\
1.48	0.003333\\
1.5	0.003009\\
1.52	0.003287\\
1.54	0.003306\\
1.56	0.003019\\
1.58	0.003276\\
1.6	0.003303\\
1.62	0.003034\\
1.64	0.003273\\
1.66	0.003026\\
1.68	0.003207\\
1.7	0.003264\\
1.72	0.003047\\
1.74	0.003263\\
1.76	0.003279\\
1.78	0.003081\\
1.8	0.003286\\
1.82	0.003159\\
1.84	0.003376\\
1.86	0.00327\\
1.88	0.003037\\
1.9	0.003299\\
1.92	0.003396\\
1.94	0.003053\\
1.96	0.003411\\
1.98	0.00308\\
};
\end{axis}

\begin{axis}[%
width=0.95092\figurewidth,
height=0.193548\figureheight,
at={(0\figurewidth,0.268817\figureheight)},
scale only axis,
xmin=0,
xmax=2,
xlabel={},
ymin=-20,
ymax=20,
ylabel={input [V]}
]
\addplot [color=black,thick,solid,forget plot]
  table[row sep=crcr]{%
0	-10.661304\\
0.02	-5.759254\\
0.04	-3.766257\\
0.06	-2.895812\\
0.08	-2.691099\\
0.1	-2.588702\\
0.12	-2.377506\\
0.14	-2.311291\\
0.16	-2.120442\\
0.18	-1.941358\\
0.2	-1.9096\\
0.22	-1.875072\\
0.24	-1.850008\\
0.26	-1.762505\\
0.28	-1.575615\\
0.3	-1.551778\\
0.32	-1.363749\\
0.34	-1.253446\\
0.36	-0.859448\\
0.38	-0.870193\\
0.4	-0.714794\\
0.42	-0.45861\\
0.44	0.709459\\
0.46	2.746722\\
0.48	5.12868\\
0.5	6.811098\\
0.52	8.630126\\
0.54	10.218879\\
0.56	11.892797\\
0.58	12.340358\\
0.6	11.445518\\
0.62	10.478223\\
0.64	8.904338\\
0.66	6.960309\\
0.68	4.072741\\
0.7	4.235745\\
0.72	3.641685\\
0.74	1.546003\\
0.76	-2.673755\\
0.78	-0.80087\\
0.8	-3.739204\\
0.82	-4.605289\\
0.84	-5.458021\\
0.86	-2.526916\\
0.88	-0.512842\\
0.9	0.872908\\
0.92	1.53862\\
0.94	0.478037\\
0.96	0.069916\\
0.98	-1.790093\\
1	-1.692252\\
1.02	-2.659752\\
1.04	-2.936021\\
1.06	-4.823726\\
1.08	-5.259986\\
1.1	-4.43912\\
1.12	-4.596825\\
1.14	-3.212773\\
1.16	-1.994402\\
1.18	-0.316775\\
1.2	-0.008817\\
1.22	0.108592\\
1.24	0.453491\\
1.26	1.649517\\
1.28	1.237875\\
1.3	1.650684\\
1.32	1.153646\\
1.34	1.704203\\
1.36	2.404551\\
1.38	2.337088\\
1.4	1.50851\\
1.42	0.448494\\
1.44	-1.223992\\
1.46	-2.527798\\
1.48	-4.425058\\
1.5	-4.832756\\
1.52	-3.679072\\
1.54	-3.247458\\
1.56	-2.174715\\
1.58	-0.684209\\
1.6	0.166364\\
1.62	0.31906\\
1.64	1.637913\\
1.66	2.547326\\
1.68	3.040443\\
1.7	3.230354\\
1.72	3.906807\\
1.74	4.322518\\
1.76	3.262485\\
1.78	1.046794\\
1.8	-0.432515\\
1.82	-1.148301\\
1.84	-1.0226\\
1.86	-1.385316\\
1.88	-2.493283\\
1.9	-2.75916\\
1.92	-2.675179\\
1.94	-3.111855\\
1.96	-2.469486\\
1.98	-1.187189\\
};
\end{axis}

\begin{axis}[%
width=0.95092\figurewidth,
height=0.193548\figureheight,
at={(0\figurewidth,0.806452\figureheight)},
scale only axis,
xmin=0,
xmax=2,
xlabel={},
ymin=-0.5,
ymax=0.5,
ylabel={cart pos. [m]}
]
\addplot [color=black,thick,solid,forget plot]
  table[row sep=crcr]{%
0	0.001026\\
0.02	-0.005805\\
0.04	-0.020107\\
0.06	-0.031519\\
0.08	-0.04252\\
0.1	-0.05524\\
0.12	-0.067344\\
0.14	-0.080763\\
0.16	-0.093676\\
0.18	-0.105957\\
0.2	-0.12033\\
0.22	-0.133792\\
0.24	-0.147374\\
0.26	-0.161277\\
0.28	-0.178542\\
0.3	-0.191273\\
0.32	-0.207256\\
0.34	-0.221633\\
0.36	-0.239587\\
0.38	-0.252129\\
0.4	-0.265726\\
0.42	-0.277891\\
0.44	-0.291437\\
0.46	-0.301213\\
0.48	-0.308844\\
0.5	-0.312054\\
0.52	-0.312157\\
0.54	-0.306169\\
0.56	-0.293322\\
0.58	-0.27383\\
0.6	-0.244809\\
0.62	-0.213874\\
0.64	-0.177356\\
0.66	-0.140292\\
0.68	-0.099303\\
0.7	-0.066436\\
0.72	-0.030498\\
0.74	0.006153\\
0.76	0.045729\\
0.78	0.073326\\
0.8	0.100766\\
0.82	0.124442\\
0.84	0.146121\\
0.86	0.15691\\
0.88	0.16601\\
0.9	0.174224\\
0.92	0.182944\\
0.94	0.190533\\
0.96	0.200073\\
0.98	0.208503\\
1	0.216991\\
1.02	0.221911\\
1.04	0.225884\\
1.06	0.22918\\
1.08	0.234236\\
1.1	0.233848\\
1.12	0.225962\\
1.14	0.214085\\
1.16	0.198931\\
1.18	0.184727\\
1.2	0.168817\\
1.22	0.154529\\
1.24	0.139339\\
1.26	0.129092\\
1.28	0.118284\\
1.3	0.110217\\
1.32	0.102913\\
1.34	0.097615\\
1.36	0.092145\\
1.38	0.086543\\
1.4	0.083969\\
1.42	0.085566\\
1.44	0.089236\\
1.46	0.092838\\
1.48	0.095435\\
1.5	0.095528\\
1.52	0.089917\\
1.54	0.079192\\
1.56	0.066434\\
1.58	0.053189\\
1.6	0.039146\\
1.62	0.025157\\
1.64	0.007631\\
1.66	0.000933\\
1.68	-0.006874\\
1.7	-0.015725\\
1.72	-0.022902\\
1.74	-0.022293\\
1.76	-0.017448\\
1.78	-0.00798\\
1.8	0.002674\\
1.82	0.009697\\
1.84	0.0168\\
1.86	0.022253\\
1.88	0.026053\\
1.9	0.028113\\
1.92	0.028881\\
1.94	0.024784\\
1.96	0.018572\\
1.98	0.012628\\
};
\addplot [color=black,dashed,forget plot]
  table[row sep=crcr]{%
0	-0.45\\
0.02	-0.45\\
0.04	-0.45\\
0.06	-0.45\\
0.08	-0.45\\
0.1	-0.45\\
0.12	-0.45\\
0.14	-0.45\\
0.16	-0.45\\
0.18	-0.45\\
0.2	-0.45\\
0.22	-0.45\\
0.24	-0.45\\
0.26	-0.45\\
0.28	-0.45\\
0.3	-0.45\\
0.32	-0.45\\
0.34	-0.45\\
0.36	-0.45\\
0.38	-0.45\\
0.4	-0.45\\
0.42	-0.45\\
0.44	-0.45\\
0.46	-0.45\\
0.48	-0.45\\
0.5	-0.45\\
0.52	-0.45\\
0.54	-0.45\\
0.56	-0.45\\
0.58	-0.45\\
0.6	-0.45\\
0.62	-0.45\\
0.64	-0.45\\
0.66	-0.45\\
0.68	-0.45\\
0.7	-0.45\\
0.72	-0.45\\
0.74	-0.45\\
0.76	-0.45\\
0.78	-0.45\\
0.8	-0.45\\
0.82	-0.45\\
0.84	-0.45\\
0.86	-0.45\\
0.88	-0.45\\
0.9	-0.45\\
0.92	-0.45\\
0.94	-0.45\\
0.96	-0.45\\
0.98	-0.45\\
1	-0.45\\
1.02	-0.45\\
1.04	-0.45\\
1.06	-0.45\\
1.08	-0.45\\
1.1	-0.45\\
1.12	-0.45\\
1.14	-0.45\\
1.16	-0.45\\
1.18	-0.45\\
1.2	-0.45\\
1.22	-0.45\\
1.24	-0.45\\
1.26	-0.45\\
1.28	-0.45\\
1.3	-0.45\\
1.32	-0.45\\
1.34	-0.45\\
1.36	-0.45\\
1.38	-0.45\\
1.4	-0.45\\
1.42	-0.45\\
1.44	-0.45\\
1.46	-0.45\\
1.48	-0.45\\
1.5	-0.45\\
1.52	-0.45\\
1.54	-0.45\\
1.56	-0.45\\
1.58	-0.45\\
1.6	-0.45\\
1.62	-0.45\\
1.64	-0.45\\
1.66	-0.45\\
1.68	-0.45\\
1.7	-0.45\\
1.72	-0.45\\
1.74	-0.45\\
1.76	-0.45\\
1.78	-0.45\\
1.8	-0.45\\
1.82	-0.45\\
1.84	-0.45\\
1.86	-0.45\\
1.88	-0.45\\
1.9	-0.45\\
1.92	-0.45\\
1.94	-0.45\\
1.96	-0.45\\
1.98	-0.45\\
};
\addplot [color=black,dashed,forget plot]
  table[row sep=crcr]{%
0	0.45\\
0.02	0.45\\
0.04	0.45\\
0.06	0.45\\
0.08	0.45\\
0.1	0.45\\
0.12	0.45\\
0.14	0.45\\
0.16	0.45\\
0.18	0.45\\
0.2	0.45\\
0.22	0.45\\
0.24	0.45\\
0.26	0.45\\
0.28	0.45\\
0.3	0.45\\
0.32	0.45\\
0.34	0.45\\
0.36	0.45\\
0.38	0.45\\
0.4	0.45\\
0.42	0.45\\
0.44	0.45\\
0.46	0.45\\
0.48	0.45\\
0.5	0.45\\
0.52	0.45\\
0.54	0.45\\
0.56	0.45\\
0.58	0.45\\
0.6	0.45\\
0.62	0.45\\
0.64	0.45\\
0.66	0.45\\
0.68	0.45\\
0.7	0.45\\
0.72	0.45\\
0.74	0.45\\
0.76	0.45\\
0.78	0.45\\
0.8	0.45\\
0.82	0.45\\
0.84	0.45\\
0.86	0.45\\
0.88	0.45\\
0.9	0.45\\
0.92	0.45\\
0.94	0.45\\
0.96	0.45\\
0.98	0.45\\
1	0.45\\
1.02	0.45\\
1.04	0.45\\
1.06	0.45\\
1.08	0.45\\
1.1	0.45\\
1.12	0.45\\
1.14	0.45\\
1.16	0.45\\
1.18	0.45\\
1.2	0.45\\
1.22	0.45\\
1.24	0.45\\
1.26	0.45\\
1.28	0.45\\
1.3	0.45\\
1.32	0.45\\
1.34	0.45\\
1.36	0.45\\
1.38	0.45\\
1.4	0.45\\
1.42	0.45\\
1.44	0.45\\
1.46	0.45\\
1.48	0.45\\
1.5	0.45\\
1.52	0.45\\
1.54	0.45\\
1.56	0.45\\
1.58	0.45\\
1.6	0.45\\
1.62	0.45\\
1.64	0.45\\
1.66	0.45\\
1.68	0.45\\
1.7	0.45\\
1.72	0.45\\
1.74	0.45\\
1.76	0.45\\
1.78	0.45\\
1.8	0.45\\
1.82	0.45\\
1.84	0.45\\
1.86	0.45\\
1.88	0.45\\
1.9	0.45\\
1.92	0.45\\
1.94	0.45\\
1.96	0.45\\
1.98	0.45\\
};
\end{axis}
\end{tikzpicture}%

%% file: conclusion.tex
\section{Conclusion}\label{Sec:Concl}
The article discussed a method for reducing the complexity of the optimization algorithms encountered in MPC by parametrizing input and state trajectories with basis functions. The basis functions are chosen to be invariant to time-shifts, which enables warm-starting of consecutive optimization problems. In addition, in the case of linear time-invariant dynamics and a quadratic cost, closed-loop stability and recursive feasibility results are obtained without the addition of a terminal set constraint and a terminal cost. The method is applied to the swing-up of an inverted pendulum system, where it is shown that the number of degrees of freedom can be greatly reduced with the parametrization. The resulting MPC algorithm runs with a sampling time of 20ms and is able to swing the inverted pendulum up while accounting for input and state constraints.

%% file: appendixNew.tex
\section{Proof of Prop.~\ref{Prop:RecFeasibilityStab}}\label{App:Proof}
We consider the optimal cost to be a function of the initial condition $x_0$,
\begin{align}
J(x_0):=&\min_{\eta_x \in \mathbb{R}^{ns}, \eta_u \in \mathbb{R}^{ms}} \sum_{k=0}^{\infty} l(\tilde{x}(k), \tilde{u}(k)) \quad \text{s.t. } \label{eq:resultingOptiApp}\\
&(I_n \otimes M\T - A \otimes I_s)\eta_x - (B \otimes I_s) \eta_u = 0,\nonumber\\
&\tilde{x}(0)=x_0,\nonumber \\
&g(\tilde{x}(k), \tilde{u}(k))\leq 0, \quad \forall k\in \{0,1,\dots,N_\text{max}\}, \nonumber
\end{align}
for all $x_0 \in X$, where $X$ denotes the set of all $x_0$ such that the above optimization is feasible.
For $x_0\in X$, the assumptions on $l$ and $g$ ($l$ is continuous, $l(0,0)=0$, and bounded below by a quadratic function, $g$ is continuous) assert that the above minimum exists and is attained, see \cite[p.~12, Ex.~1.11]{Rockafellar}. They further imply that $J$ is lower bounded,
\begin{equation}
J(x_0) \geq l(x_0,\tilde{u}(0)) \geq \underline{\sigma} |x_0|^2, \quad \forall x_0\in X.
\end{equation}
%We consider the auxilary problem (where the inequality constraints have been removed) and define
%\begin{align}\label{eq:auxProb}
%&\hat{J}(x_0):=\min \bar{\sigma} \left( \eta_x\T (I_n \otimes J_s) \eta_x + \eta_u\T (I_m \otimes J_s) \eta_u \right) \\
%&\text{s.t. } (I_n \otimes M\T - A \otimes I_s)\eta_x - (B \otimes I_s) \eta_u = 0,\nonumber\\
%&\hspace{15pt} (I_n \otimes \tau(0))\T \eta_x=x_0,\nonumber
%\end{align}
%for all $x_0 \in X$. We note that for any $\eta_x\in \mathbb{R}^{ns}$ and $\eta_u \in \mathbb{R}^{ms}$
%\begin{equation}
%\sum_{k=0}^{\infty} l(\tilde{x}(k),\tilde{u}(k))\leq \bar{\sigma} \left(\eta_x\T (I_n \otimes J_s)\eta_x + \eta_u\T (I_n \otimes J_s)\eta_u\right). \label{eq:upperBound}
%\end{equation}
Due to the fact that the inequality constraints in \eqref{eq:resultingOptiApp} describe a closed set that includes $\eta_x=0$, $\eta_u=0$ in its interior ($g$ is assumed to be continuous and $g(0,0)<0$), the regularity of equality constraints and the quadratic upper bound on the running cost imply that the inequality constraints are not active for small enough $x_0$. According to \cite[p.~16, Thm.1.17]{Rockafellar}, $J(x_0)$ is therefore continuous at zero. 

We now consider the trajectory of the closed-loop system $x(k)$, $k\in \mathbb{Z}^+$ starting at any $x_0\in X$. According to Prop.~\ref{Prop:LinDynamics}, the trajectories $\tilde{x}(k)$ and $\tilde{u}(k)$ resulting from solving the optimization \eqref{eq:resultingOptiApp} at time $0$ satisfy the dynamics and the constraints for all times. Due to the time-shift property of the basis functions, the trajectories $\tilde{x}(k+1)$ and $\tilde{u}(k+1)$ lie likewise in the span of the basis functions, and therefore, they represent feasible candidates for the optimization at the next time step, i.e. for $k=1$. Moreover, the optimal cost at the next time step ($k=1$), $J(x(1))$, is therefore bounded by
\begin{equation}
J(x(1))\leq \sum_{k=0}^{\infty} l(\tilde{x}(k+1),\tilde{u}(k+1)).
\end{equation}
The right-hand side can be expressed in terms of the cost $J(x(0))$, that is,
\begin{equation*}
\sum_{k=0}^\infty l(\tilde{x}(k),\tilde{u}(k)) - l(\tilde{x}(0),\tilde{u}(0))=J(x(0))-l(\tilde{x}(0),\tilde{u}(0)),
\end{equation*}
which yields, (using $\tilde{x}(0)=x(0)$ and $\tilde{u}(0)=u(0)$),
\begin{equation}
J(x(1))-J(x(0)) \leq -l(x(0),u(0))
\end{equation}
It follows by induction that \eqref{eq:resultingOptiApp} is recursively feasible, and that, for any $k\geq 0$,
\begin{equation}
J(x(k+1))-J(x(k)) \leq -l(x(k),u(k)) \leq -\underline{\sigma} |x(k)|^2.\label{eq:decayTmp}
\end{equation}
Hence, the cost $J(x(k))$ forms a monotonically decreasing sequence (in $k\in \mathbb{Z}^+$) that is bounded below by 0 and therefore converges. In the limit as $k\rightarrow \infty$, \eqref{eq:decayTmp} reduces to
\begin{equation}
0\leq \lim_{k\rightarrow\infty} -\underline{\sigma} |x(k)|^2,
\end{equation}
implying that $x(k) \rightarrow 0$ as $k\rightarrow \infty$. This shows that the origin is attractive in $X$.

It remains to show that the origin is stable in the sense of Lyapunov. Note that attractivity alone does not imply stability. For proving stability we pick any $\epsilon >0$. Due to the continuity of $J(x_0)$, there exists a $\delta > 0$, such that $|x_0|<\delta$ implies $J(x_0)<\underline{\sigma} \epsilon^2$ (and $x_0 \in X$). As argued in the previous paragraph, the cost $J(x(k))$ of any closed-loop trajectory $x(k)$ starting at a $|x_0|<\delta$ is monotonically decreasing. This concludes that $\underline{\sigma} |x(k)|^2 \leq J(x(k)) \leq J(x_0) < \underline{\sigma} \epsilon^2$, that is, $|x(k)|< \epsilon$.\hfill\qed

\subsection*{\HChange{Remarks}}
\HChange{The section concludes with the two remarks that indicate further extensions and generalization of the recursive feasibility and closed-loop stability results:}
\begin{itemize}
\item[1)] \HChange{The results can be extended to a running cost that includes a discount, that is,}
\begin{equation}
l(k,x,u)=\rho_x^k x\T Q x+ \rho_u^k u\T R u,
\end{equation}
\HChange{where $\rho_x, \rho_u \in \mathbb{R}$ are in the interval $(0,1)$, and $Q\in \mathbb{R}^{n\times n}$ is positive definite and $R\in \mathbb{R}^{m\times m}$ is positive semi-definite.}

\HChange{The results can also be extended to the trajectory-tracking case. More precisely, provided that the dynamics are linear and time-invariant, and that the reference input and state trajectories are spanned by the basis functions, the arguments of Prop.~\ref{Prop:RecFeasibilityStab} can be used to conclude recursive feasibility and asymptotic stability of the tracking error.}
\item[2)] \HChange{If the dynamics are nonlinear, recursive feasibility and closed-loop stability are no longer inherent to the problem formulation. However, our approach could be modified to enforce the nonlinear dynamics exactly (for example by eliminating the state trajectory) and to include similar stabilizing terminal conditions as in a finite-horizon non-parametrized approach, see e.g.~\cite[Ch.~5]{GrueneNonlinear}. In that case, the arguments presented in \cite[Ch.~5]{GrueneNonlinear} would apply with minor modifications. However, the computational complexity of the resulting MPC algorithm would most likely increase. In all these stability arguments, the fact that the parametrization is invariant to time-shifts is of paramount importance.}
\end{itemize}

\section{Experimental results}\label{App:Exp}

\subsection{Hardware setup and first-principles model}
\begin{figure}
\center
\includegraphics[scale=.06]{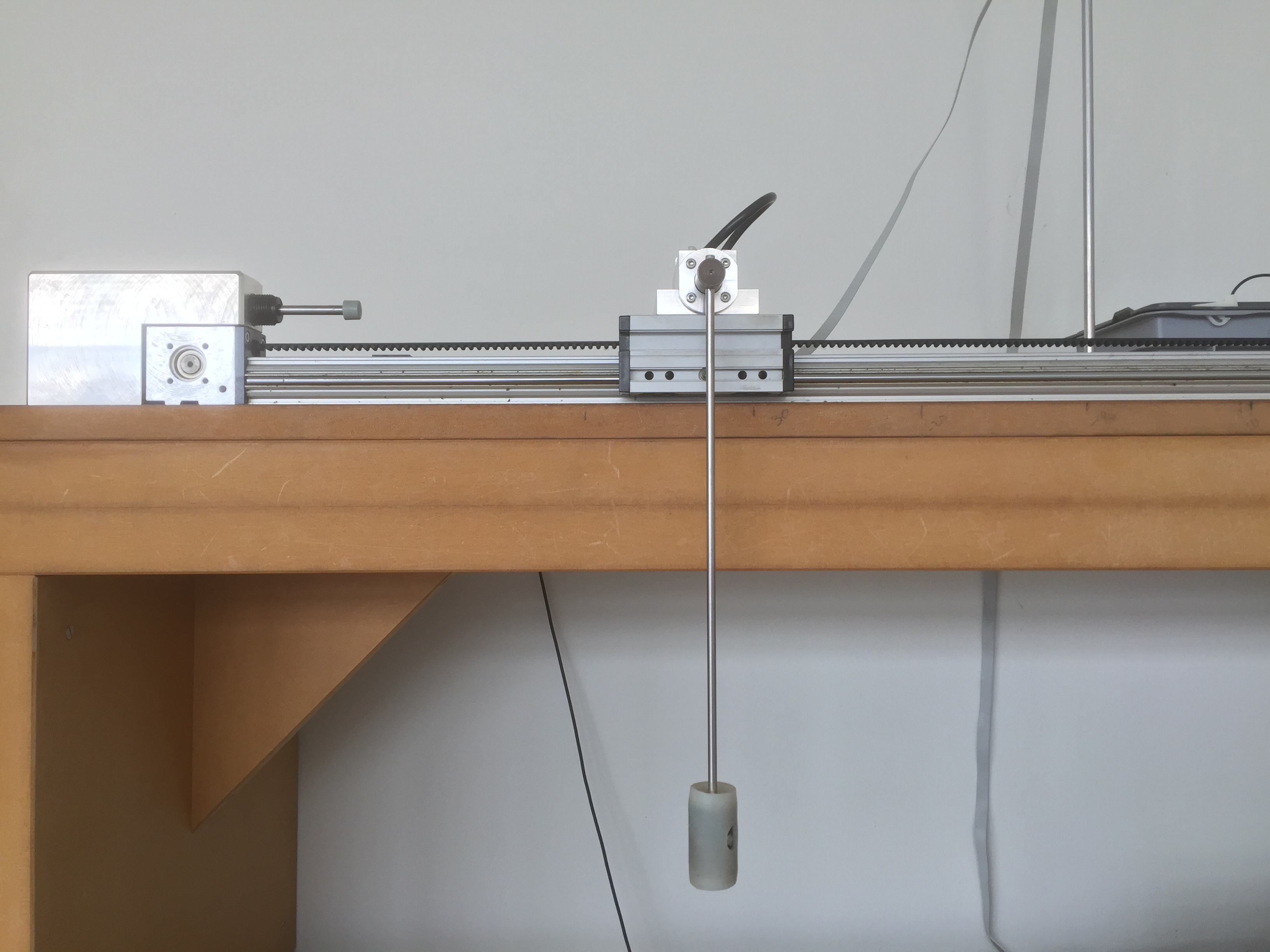}
\caption{The pendulum-on-a-cart system.}
\label{Fig:Pend}
\end{figure}

Fig.~\ref{Fig:Pend} shows the experimental setup. 
The cart is actuated by an electrical motor via a transmission belt. The voltage that can be applied to the electrical motor represents the input $u$ to the system and is limited to $\pm24$V. The cart is attached to a rail of 0.9m length. 
The position of the cart and the angle of pendulum are measured with encoders. An extended Kalman filter is used to estimate the state $x:=(x_\text{c},\dot{x}_\text{c},\varphi,\dot{\varphi})$. The control and estimation algorithms are run on a laptop (Intel Core i7, 2.6GHz, 8GB random access memory) that is interfacing the pendulum system through a serial communication. All the control and estimation algorithms run at 50Hz.

The following first-principles model is used
\begin{align}
\ddot{x}_\text{c}&=\frac{\frac{F_\text{u}}{m}-g \sin(\varphi) \cos(\varphi) + l \dot{\varphi}^2 \sin(\varphi)}{\frac{M}{m}+ \sin(\varphi)^2},\\
\ddot{\varphi}&=\frac{-\frac{F_\text{u}}{ml} \cos(\varphi)+ \frac{M+m}{ml} g \sin(\varphi) - \dot{\varphi}^2 \sin(\varphi) \cos(\varphi)}{ \frac{M}{m}+ \sin(\varphi)^2},\nonumber\\
F_\text{u}&=\frac{k_\text{m}}{r_\text{zr} R_\text{m}} (u- \frac{\dot{x}_\text{c}}{k_\text{n} r_\text{zr}}),\nonumber
\end{align}
where $F_\text{u}$ denotes the force driving the pendulum, $x_\text{c}\in [-0.45$m$,0.45$m$]$ the cart position, and $\varphi$ the pendulum angle ($\varphi=0$ corresponds to the upright equilibrium). The parameters are listed in Tab.~\ref{Tab:Params}. Friction is neglected in the model used for the MPC controller, but dry friction between the cart and rail is compensated with feedforward. The parameters of the first-principles model are identified by fitting different step-responses about the lower equilibrium.

\begin{table}
\center
\begin{tabular}{l|l|l}
\hline
parameter &value &description \\ \hline 
$m$		& 0.17kg & pendulum mass\\
$M$ 	& 0.74kg & cart mass (lumped)\\
$l$		& 0.30m	 & pendulum length\\
$k_\text{m}$ & 0.011Nm/A & torque constant\\
$k_\text{n}$ & 20.62V/s & speed constant\\
$R_\text{m}$ & 0.30$\Omega$ & resistance\\
$r_\text{zr}$ & 0.018m & belt wheel radius\\ \hline
\end{tabular}
\caption{Parameters of the first-principle model of the pendulum-on-a-cart system.}
\label{Tab:Params}
\end{table}

\subsection{\HChange{Robustness with respect to the initial guess}}
\begin{figure}
\setlength{\figurewidth}{.8\columnwidth}
\setlength{\figureheight}{.4\columnwidth}
\input{img/robustness.tikz}
\caption{\HChange{The plot shows how the closed-loop cost varies as a function of the disturbance amplitude perturbing the state at each time step.}}
\label{Fig:rob}
\end{figure}
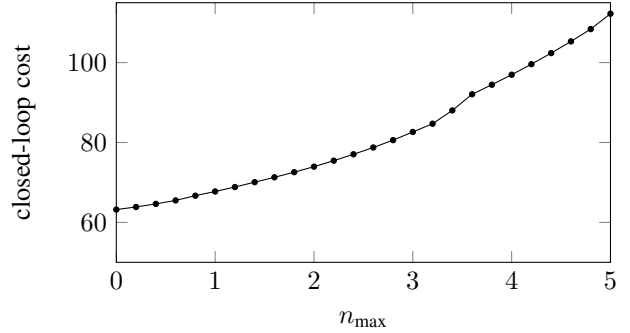

\HChange{The robustness with respect to the initial guess that starts the optimization at each time step is studied in simulations. In general, an initial guess that is close to a successful swing-up trajectory is required to start the optimization (such as the one given by Fig.~\ref{Fig:InitialGuess}). In order to quantify the robustness of the optimization with respect to the initial guess, we included the following artificial disturbance of the state variable, while performing (closed-loop) swing-up maneuvers in simulation,
\begin{equation}
x(k+1)=f(x(k),u(k))+ \left(\begin{array}{c} 0.02\text{m} \\ 0.005\text{m/s} \\ 1\text{deg} \\ 0.5\text{deg/s} \end{array}\right) n(k),
\end{equation}
where $f(x(k),u(k))$ are the nominal dynamics and $n(k)$ is a disturbance, uniformly distributed in $[-n_\text{max},n_\text{max}]$ and independent across time. The optimization is thus warm-started with the solution obtained from the previous time step, but has no information about the perturbation $n(k)$.
The resulting closed-loop cost as a function of $n_\text{max}$ is shown in Fig.~\ref{Fig:rob}.
Even with a value of $n_\text{max}$ of $5$, which amounts to disturbances of the cart of $\pm 0.1\text{m}$ and disturbances of the pendulum angle of $\pm 5\text{deg}$ at each time step, the optimization algorithm still converges at each time step, resulting in successful (closed-loop) swing-up trajectories. The effect of the disturbances is clearly visible in the closed-loop cost, which increases by almost a factor of 2. Increasing the value of $n_\text{max}$ further leads to violations of the position constraint due to the limited rail length.}

\HChange{In our experience, it is important that the optimization is able to handle these disturbances well, as these also occur in the experimental setup due to model mismatch.}

%% file: img/robustness.tikz
% Minimal pgfplots version: 1.3
%  http://www.mathworks.com/matlabcentral/fileexchange/22022-matlab2tikz
\begin{tikzpicture}

\begin{axis}[%
width=0.95092\figurewidth,
height=\figureheight,
at={(0\figurewidth,0\figureheight)},
scale only axis,
xmin=0,
xmax=5,
xlabel={$n_\text{max}$},
ymin=50,
ymax=115,
ylabel={closed-loop cost},
legend style={legend cell align=left,align=left,draw=white!15!black}
]
\addplot [color=black,solid,mark=*,mark size=1pt,mark options={solid},forget plot]
  table[row sep=crcr]{%
0	63.2075279510331\\
0.2	63.8604113675944\\
0.4	64.6294119554752\\
0.6	65.4998283657576\\
0.8	66.6877473970442\\
1	67.7372185488921\\
1.2	68.8497552081575\\
1.4	70.0741763088164\\
1.6	71.2909793103555\\
1.8	72.5691175044464\\
2	73.9511719132007\\
2.2	75.4504279824796\\
2.4	77.0522971826367\\
2.6	78.7535061256339\\
2.8	80.5896863495487\\
3	82.643775560011\\
3.2	84.7037908899148\\
3.4	88.0357707027977\\
3.6	92.0830884426417\\
3.8	94.4702261728071\\
4	96.9728012521211\\
4.2	99.6095156107587\\
4.4	102.386593911013\\
4.6	105.312753979828\\
4.8	108.392148250299\\
5	112.254483621373\\
};
\end{axis}
\end{tikzpicture}%